\documentclass[prl,twocolumn,preprintnumbers,footinbib,tightenlines,superscriptaddress]{revtex4-1}
\usepackage[utf8]{inputenc}
\usepackage{graphicx,amsmath,amsfonts,amssymb,aas_macros,slashed}
\usepackage{datetime}
\usepackage{numprint}
\usepackage{enumitem}
\usepackage{booktabs}
\usepackage{numprint}
\usepackage{bbold}
\usepackage{mathrsfs}
\usepackage{float}

\usepackage{graphicx,color,amsmath}
\usepackage{hyperref}

\usepackage{cleveref}
\crefname{equation}{Eq.}{Eqs.} 

\allowdisplaybreaks

\DeclareMathOperator{\Tr}{Tr}

\def\figureautorefname~#1\null{Fig.\,#1\null}
\def\equationautorefname~#1\null{Eq.\,(#1)\null}
\def\tableautorefname~#1\null{Tab.\,#1\null}

\usepackage{pgfplots}
\usepackage{bm}
\usepackage{mathtools}
\usepackage{braket}

\begin{document}

\title{\bf Kinetic Equilibrium between SIMP Dark Matter and Radiation via Internal Bremsstrahlung}

\author{Xiaoyong Chu}
\email{chuxiaoyong@ucas.ac.cn}
\affiliation{International Centre for Theoretical Physics Asia-Pacific, University of Chinese Academy of Sciences, 100190 Beĳing,China}

\author{Ajay Kaladharan}
\email{kaladharan.ajay@ucas.ac.cn}
\affiliation{International Centre for Theoretical Physics Asia-Pacific, University of Chinese Academy of Sciences, 100190 Beĳing,China}

\author{Yassine El Mohtadi}
\email{yassou.kidou@mails.ucas.ac.cn}
\affiliation{International Centre for Theoretical Physics Asia-Pacific, University of Chinese Academy of Sciences, 100190 Beĳing,China}

\begin{abstract}

As an alternative to standard dark matter freeze-out via two-body annihilation, strongly interacting massive particles are proposed to regulate their relic abundance through three/four-body self-annihilation. During such freeze-out, dark matter must undergo both number reduction and energy dissipation. The former requires number-reducing processes, and the latter requires an efficient energy-dissipation mechanism. In this work, we propose a novel mechanism where internal Bremsstrahlung processes establish kinetic equilibrium between dark matter and radiation, and thus dissipate energy. More appealingly, this mechanism can be naturally realized in a class of strong dynamics models with flavor-universal quarks and a light axion. To demonstrate our results in a $Sp(4)_c$ dark-color model, we compute all relevant interaction terms and numerically solve the full Boltzmann equations. The same mechanism can also be applied to vector dark matter models.

\end{abstract}

\maketitle

\begin{sloppypar}


{\bf Introduction.\,}While the existence of astrophysical dark matter (DM) is well established, and its fractional contribution to the present cosmic energy budget has been precisely measured~\cite{Planck:2018vyg}, how it obtained its observed relic density is still an open question. Under the assumption that  DM particles were initially well populated, they have to go through sufficient annihilation to reduce their co-moving number and energy densities  in the early Universe, usually referred to freeze-out~\cite{Zeldovic:1965rys,Lee:1977ua,Gondolo:1990dk,Kolb:1990vq}. Conventionally, it is  thought that a pair of DM particles should annihilate into lighter species. Nevertheless, in the case that the DM candidate is a Strongly Interacting Massive Particle (SIMP), freeze-out via self-annihilation, such as $3\pi\to 2\pi$ processes, is also possible~\cite{Carlson:1992fn,deLaix:1995vi}.  Successful SIMP freeze-out was proposed in~\cite{Hochberg:2014dra}, and has since been extensively studied; see, e.g.,~\cite{Hochberg:2014kqa,Hochberg:2015vrg,Braat:2023fhn,Kulkarni:2022bvh,Chu:2024rrv,Kuflik:2015isi,Bernal:2015bla, Bernal:2015xba, Bernal:2015ova,  Choi:2015bya, Choi:2016hid,Soni:2016gzf, Kamada:2016ois,  Bernal:2017mqb,Cline:2017tka, Choi:2017mkk,  Kuflik:2017iqs, Heikinheimo:2018esa, Choi:2018iit,Hochberg:2018rjs, Bernal:2019uqr, Choi:2019zeb,  Katz:2020ywn, Smirnov:2020zwf, Xing:2021pkb,Bernreuther:2023kcg,Garcia-Cely:2024ivo,Alfano:2025non,Garcia-Cely:2025flv,Fiorentino:2026dea,Kamada:2022zwb}.  We will use $\pi$ to denote the  DM particle, although SIMP candidates are not limited to dark pion models.  

One essence of the SIMP freeze-out scenario~\cite{Hochberg:2014dra,Hochberg:2014kqa} is the kinetic equilibrium between DM particles and a radiation bath, to ensure that DM particles can dissipate their kinetic energy efficiently and to suppress the inverse process of self-annihilation. This   equilibrium is typically maintained by DM-radiation elastic scattering. In such a set-up, crossing symmetry relates DM-radiation  scattering and DM pair annihilation, allowing for sizable DM annihilating into radiation. This usually leads  to very strong constraints on the allowed parameter space. In this letter, we explore an exception beyond this crossing symmetry argument, where the scattering and annihilation between DM and radiation are characterized by different combinations of parameters. This thus allows for extra freedom for SIMP model building.    
 
\begin{figure}[t]
    \centering
    \includegraphics[width=0.38\textwidth]{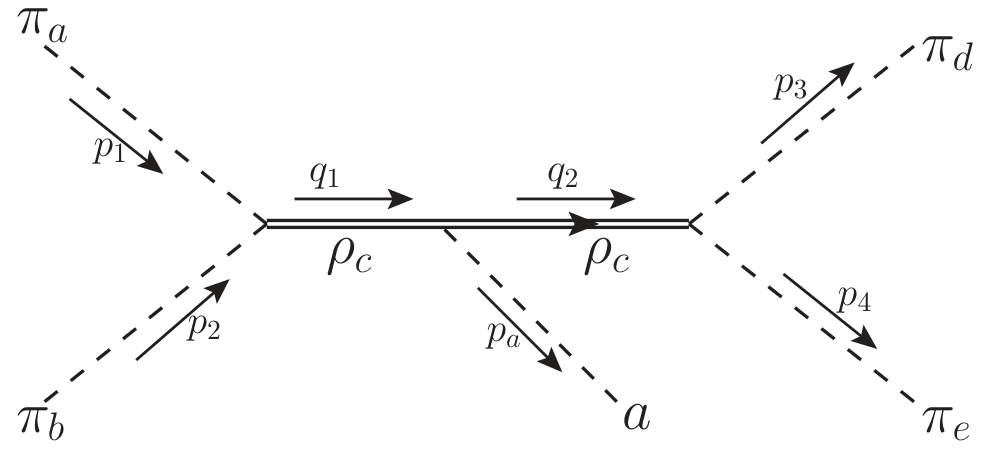}
    \caption{Process of $\pi_a\, \pi_b \leftrightarrow \pi_d\, \pi_e \,a$, which maintains the kinetic equilibrium between DM particles, $\pi$, and dark radiation, $a$. This is mediated by a (off-shell)  heavy particle,  $\rho$.}
    \label{fig:inter}\vspace{-0.4cm}
\end{figure}

Concretely, we suggest that kinetic equilibrium can be realized instead by internal Bremsstrahlung processes, illustrated by~\autoref{fig:inter}:  
\begin{equation}\label{eq:processDB}
 \pi +  \pi \,\leftrightarrow\,    \pi  + \pi + {\mathcal {R}}\,,
\end{equation}
where ${\mathcal {R}}$ stands for a  nearly massless dark particle, i.e., dark radiation during  freeze-out, which only directly couples to the intermediate particles. In practice, such processes can have many realizations, such as an intermediate particle $X$, whose mass is set to be larger than $2m_\pi$, via 
\begin{equation}\label{eq:processSemi}
\pi +  \pi \,\leftrightarrow\,  X + {\mathcal {R}}\,, 	
\end{equation}
where $X$ subsequently decays into two pions either on-shell or off-shell. If $X$  appears on-shell, the latter process,   \autoref{eq:processSemi}, should be treated as the resonant part of internal Bremsstrahlung given in  \autoref{eq:processDB}. 
{Unlike the previous SIMP set-up, in which kinetic equilibrium is maintained through DM–$\mathcal{R}$ elastic scattering, our mechanism does not have a direct DM–$\mathcal{R}$ coupling.} As a result, DM two-body annihilation, such as $\pi +  \pi\,\leftrightarrow\, \pi  + {\mathcal {R}} $, can be greatly suppressed. 
Moreover, by making the intermediate particle, $X$, couple to the DM particle much more strongly than to the radiation,  ${\mathcal {R}}$, one has processes that involve two-or-more ${\mathcal {R}}$ particles  effectively decoupled. Therefore, $X$-induced $2\pi \to 2{\mathcal {R}}$ annihilation can be neglected. This also guarantees that during DM freeze-out $3\pi \to 2\pi$ self-annihilation dominates over other annihilation processes like $3\pi \to \pi +{\mathcal {R}}$ and $\pi +X \to \pi +{\mathcal {R}}$. The suppression of the last process relies on the fact that during freeze-out the abundance of $X$ is much smaller than that of the DM particles for $m_X \gtrsim  2m_\pi$~\footnote{In the case of $m_X < 2m_\pi$, such as $X$ being a two-$\pi$ bound state, the abundance of $X$ may become much larger than its thermal equilibrium value. This could strongly enhance DM freeze-out; see Ref.~\cite{Chu:2024rrv} for an example.}.

Remarkably, the requirements above are naturally realized in a class of dark pion DM candidates. In this class of models, the intermediate particle,  $X$, is a dark meson, such as $\rho$, slightly heavier than dark pions, while the role of  ${\mathcal {R}}$ is played by a dark axion, after introducing  a dark $U(1)$ Peccei-Quinn (PQ) symmetry\,\footnote{Axion-induced kinetic equilibrium has been considered for the SIMP models, but via elastic scattering; see e.g.  \cite{Hochberg:2018rjs,Fiorentino:2026dea}. }. We show that one can easily forbid the direct coupling between dark pions and the dark axion, as well as suppress the coupling between the $\rho$ meson and the axion. They are achieved by imposing  flavor universality for   dark quarks, as well as  a hierarchy between the dark pion decay constant and  PQ breaking scale, $f_\pi \ll f_a$.  Below we adopt an exemplary $Sp(N_c)$ gauge theory  to introduce these relevant interaction terms following Ref.~\cite{Bai:2024lpq,Bai:2025fvl},  and to demonstrate how SIMP freeze-out can happen successfully by imposing  the two conditions above, flavor universality and scale hierarchy. More details, including the calculation of effective interaction terms, thermally-averaged interaction rates, and the full evolution Boltzmann equations,  are provided in the supplementary material.

{\bf Dark QCD model with an axion.}\, 
We consider a $Sp(N_c)$ gauge theory with the color $N_c=4$  for two Dirac quarks, $u$ and $d$, in the dark sector. As  mentioned above, we set this theory to be flavor universal, so $m_u =m_d$.  As the quarks are described by a pseudo-real representation of the color group, the corresponding global flavor symmetry of the four Weyl spinors is actually $SU(4)$.   The fermion mass term and color confinement break $SU(4)$ to $Sp(4)$, yielding five equally massive dark pions, which serve as the DM candidate of the theory. 
Due to the non-trivial fifth homotopy group of the coset space $SU(4)/Sp(4)$,  the Wess–Zumino–Witten (WZW) term generates a five-pion interaction that naturally mediates the $3\pi\to 2\pi$ process~\cite{Wess:1971yu, Witten:1983tw} via 
\begin{equation}
\Gamma_{\mathrm{WZW}}^{0}=-\frac{iN_c}{240\pi^2}\int \mathrm{Tr} \left( \Sigma^\dagger d \Sigma \right)^5 \,, \label{eq:WZW_L}
\end{equation}
in which the pion fields are parametrized as $\Sigma=\exp\left({i\pi /f_\pi }\right)\Sigma_0 \exp\left(i\pi^T /f_\pi  \right)$, with $\pi = \sum^5_{a=1} \pi_a T^a$. Here    $\Sigma_0=i \sigma_2 \otimes \mathcal{I}_{2}$ is vacuum condensate, $f_\pi$ is the pion decay constant, and $T_a$ are generators of the    coset space $SU(4)/Sp(4)$.   
For the standard SIMP models~\cite{Hochberg:2014dra, Hochberg:2015vrg}, this interaction terms generate the observed DM relic abundance through $3 \pi \to 2\pi $  self-annihilation. The strong dynamics of this framework has also been extensively studied using lattice simulations~\cite{TELOS:2026alk,Pica:2016zst,Bennett:2017kga,Bennett:2019jzz,Bennett:2019cxd,Bennett:2022yfa,Bennett:2023wjw}. 

Now we add a $SU(4)$-singlet  axion to the model. Before color confinement, the effective Lagrangian terms involving the axion has the form of 
\begin{align}
\mathcal{L}_{\rm{eff}}& \supset  
\frac12 \partial_\mu a \partial^\mu a -\frac{m^2_{a,0}}{2}a^2  + c_{\mathrm{gg}}\frac{\alpha}{4\pi}\frac{a}{f_a}G_{\mu \nu}\tilde{G}^{\mu\nu} \nonumber \\
&+ \frac{\partial_\mu a}{f_a}\left(\mathbb{k}_{L,0} \bar{q}_L\gamma^\mu q_L+\mathbb{k}_{R,0} \bar{q}_R\gamma^\mu q_R  \right) 
\, ,  
 \label{eq:quarkeff1}
\end{align}
where $q=(u,\,d)^T$, $f_a$ is the PQ breaking scale, and $c_{gg}$ characterizes the anomalous coupling between the axion and the dark color strength tensor $G_{\mu\nu}$, with $\tilde G^{\mu\nu} \equiv {1\over 2}\epsilon^{\mu\nu \lambda  \sigma} G_{\lambda  \sigma} $\,\footnote{In our model there is no other gauge symmetry except for $Sp(N_c)$, so we have neither new fundamental gauge fields $\mathbb{A}$ nor their associated counter-terms,  which are used to maintain new gauge symmetries. }. 
In addition, $\mathbb{k}_{L/R,0}$ denotes the dark quark-axion coupling, where flavor universality enforces identical axion couplings for the two quarks, in matrix form of $\mathbb{k}_{L/R,0}=\rm{diag}(k_{L/R, 0},\, k_{L/R, 0})$. By setting $k_{L,0}=-k_{R,0}$, a global $SU(4)$ flavor symmetry among the four Weyl spinors, is still conserved.  This is transparent in the Nambu–Gorkov formalism, which introduces a quadruplet $\Psi\equiv (u_L,\,d_L,\,\sigma_2 S u_R^\star,\,\sigma_2 S d_R^\star)^T$, satisfying $S^2=-\mathbb{1}$ and $(\tau_a)^T=S\tau_a S$ for $\tau_a\in Sp(N_c)$ generators. For details, see  Ref.~\cite{Kulkarni:2022bvh, PhysRev.117.648, osti_7264935,Kogut:2000ek} and our supplementary material.

To match the standard chiral Lagrangian, following the previous work\,\cite{Bauer:2021wjo, Bai:2024lpq}, we absorb the $aG\tilde{G}$ term via an axion-dependent phase rotation of each quark field:
\begin{equation}
q(x) \rightarrow \mathrm{exp}\left[ -i \mathbb{\kappa}_{q,0}\gamma_5  c_{gg} \frac{a(x)}{f} \right]q(x)\, ,
\label{eq:chiral_rot1}  
\end{equation}
where  $\mathbb{\kappa}_{q,0}$ is a diagonal matrix with equal entries with $\mathrm{Tr}(\mathbb{\kappa}_{q,0})=1$. The $aG\tilde{G}$ term vanishes after the rotation. In the chiral representation, the same rotation can be expressed as $\mathbb{\theta}_{L/R,0} = \mp \mathbb{\kappa}_{q,0}$. 
For our flavor-universal case, the new  coefficient matrix  $\mathbb{k}_{L} = -\mathbb{k}_{R}  =  (\mathbb{k}_{L,0} +c_{gg} \mathbb{\theta}_{L,0})$,  
in practice independent of the axion field. That is, the global $SU(4)$ symmetry remains explicit after the rotation.

In this set-up, we assume that the vector meson, $\rho$, is the lightest excited meson,  which serves as the intermediate particle $X$. Its interaction with dark pions is  incorporated by treating them as background fields $\mathbb{B}$\,\footnote{ Above the confinement scale, the $\mathbb{B}$ fields are considered as fictitious background gauge fields tracking global-global-gauge and global-global-global anomalies.}, which can be written in a differential one-form as  
\begin{equation}
\mathbb{B} = \mathbb{B}^\mu dx_\mu = g_\rho \sum_{n=6}^{15} \rho^\mu_n T^n dx_\mu\,,
\end{equation}
corresponding to vector mesons associated with the unbroken generators of $SU(4)$.  
We do not include (axial-) vector mesons, which are even  heavier.
After confinement, their relevant interactions are embedded in the chiral Lagrangian terms\,
\begin{equation}
\mathscr{L}_{\chi\mathrm{PT}} \supset  \frac{f^2_\pi}{4}\mathrm{Tr}\left[ (D  \Sigma )(D \Sigma)^\dagger \right] +  c_{\rho \rho} \frac{a}{f_a} F_{\mu \nu}(\rho)\tilde{F}^{\mu \nu}(\rho)\,,
\end{equation}
as well as from the gauged  WZW action and counter terms~\cite{Harvey:2007ca}.  
In addition, associated axion interactions is also incorporated into the background fields~\cite{Son:2004tq, Bauer:2021wjo,Bai:2023bbg} via a replacement 
\begin{equation}
\mathbb{B}\to\mathbb{B}+\mathbb{K}(a)\frac{da}{f_a}\,,
\end{equation}
where $\mathbb{K}(a)=\mathrm{diag}\left\{ \mathbb{k}_L, -\mathbb{k}_R \right\}\equiv \mathsf{K}\cdot\mathbb{1}_4$. 
Moreover, the interactions among mesons and the axion are also encoded in the covariant derivative 
\begin{equation}
D_\mu \Sigma \approx \partial_\mu \Sigma+ig_\rho\left( \rho_\mu \Sigma+\Sigma\rho_\mu^T \right)+i\frac{\partial_\mu a}{f_a}\left(\mathbb{K} \Sigma+\Sigma \mathbb{K}^T  \right) .
\end{equation}
Now the obtained meson-axion Lagrangian preserves a global $Sp(4)$ symmetry with flavor universality and $k_{L, 0}= - k_{R, 0}$. 
For numerical results below we will set the effective coupling $g_\rho  =  5$,   $c_{\rho \rho} =\frac 12$, and $\mathsf{K}=- \frac 14$.
 

{\bf Relevant interaction terms.\,}
This section  collects the relevant interaction terms in determining the evolution of the dark pion abundances. As usual, the DM number is changed by the $3\pi\to 2\pi$ self-annihilation process, which has been well calculated in the literature~\cite{Hochberg:2014kqa, Kamada:2022zwb}. Here we focus on the internal Bremsstrahlung process,  $2\pi \leftrightarrow 2\pi+a$,  in this framework, through which the  kinetic equilibrium between dark pions and radiation is maintained, as illustrated in \autoref{fig:inter}.

This internal Bremsstrahlung is mediated by the $\rho$ meson, for which  the $\pi$-$\pi$-$\rho$ vertex arises from the kinetic term in the chiral Lagrangian 
\begin{equation}
\frac{f^2_\pi}{4}\mathrm{Tr}\left[ ( D_\mu \Sigma  ) ( D^\mu \Sigma )^\dagger  \right]\supset 2i g_\rho \mathrm{Tr}\left[ \partial_\mu \pi\left( \rho^\mu\pi-\pi \rho^\mu \right) \right]\,.\notag 
\end{equation}
Meanwhile, the $a$-$\rho$-$\rho$ vertex comes from  both the chiral Lagrangian 
\begin{equation}
c_{\rho \rho}\frac{a}{f_a} \sum_{n,m}  F_{\mu \nu}^n(\rho) \tilde{F}^{m, \mu \nu} (\rho)\supset \frac{g_\rho^2   c_{\rho \rho}\epsilon^{\mu \nu \lambda \sigma} }{f_a}  a  \partial_\mu \rho^n_{\nu} \, \partial_\lambda \rho^n_{\sigma} \,,\notag 
\end{equation}
and the gauged WZW action~\cite{Duan:2000dy, Molinaro:2017mwb} 
\begin{align}
10i\mathscr{C}\mathrm{Tr}[ \left( d\mathbb{B}\mathbb{B}  +\mathbb{B}d\mathbb{B} \right) \Sigma \mathbb{B}^T \Sigma^\dagger]  \supset   \frac{20i\mathscr{C} g_\rho^2 \mathsf{K}\epsilon^{\mu \nu \lambda \sigma} }{f_a}  a  \partial_\mu \rho_\nu^n \partial_\lambda \rho_\sigma^n  \,,\notag 
\end{align}
where $\mathscr{C}=iN_c/(240\pi^2)$. The Greek letters indicates Lorentz indices and $n$ sums up all ten $\rho$ mesons ($n =6,\,...,\,15$). 
 The contribution from the WZW action is typically subleading. 
We refer to the supplementary material for detailed calculations.

One might expect $\pi$-$a$ mixing to arise from both the kinetic and mass terms in the Lagrangian; however, it vanishes because the axion couples identically to all quark flavors. Likewise, the $ \pi^3 a$ interactions from these terms vanish. From the gauged WZW action,  the $ \pi^3 a$ term also vanishes owing to the derivative nature of the axion coupling. 
These can  be understood from symmetry arguments, as the five pions are known to be in the adjoint antisymmetric representation of an  unbroken global $Sp(4)$\,\cite{Bijnens:2011fm, Chu:2024rrv}, $\bf{5}$, while the axion is a $Sp(4)$ singlet.  As mentioned above, this axion does not break  $Sp(4)$; therefore, a  $\pi$-$a$ mixing is not allowed. Similarly, the tensor product $\bf{5}\otimes \bf{5} \otimes \bf{5}$ does not yield a singlet~\cite{TELOS:2026mim,DiLuzio:2022tbb}, so   $\pi^3 a $ terms are forbidden too. While the symmetric trace component of $\bf{5}\otimes \bf{5} $ is a singlet, such a $\pi^2 a$ term necessarily violates parity, thus can only arise from the WZW term.  However, the WZW Lagrangian is a four-form and does not accommodate such three-form terms. 
Besides, we do not further consider interactions that contain two or more axions, which are doubly suppressed by the PQ scale $f_a$. 


\begin{figure}[t]
\begin{center}
\includegraphics[width=0.49\textwidth]{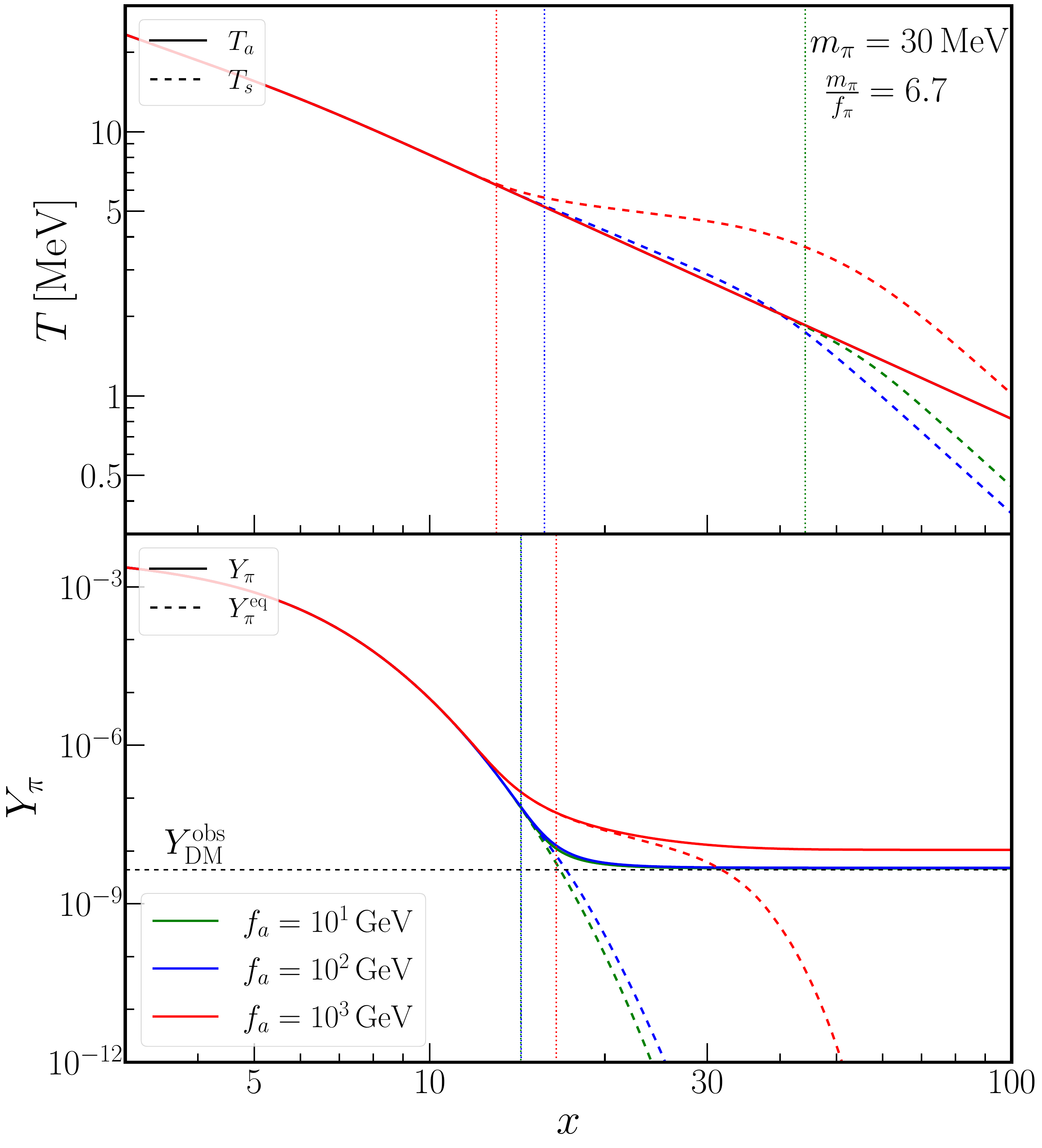}
\end{center}
\caption{Evolution of the relevant temperatures (upper panel) and comoving number density $Y_\pi$ (lower panel) for $f_a=10^2\,\mathrm{GeV}$, $m_\pi=10^2\,\mathrm{GeV}$, and $m_\pi/f_\pi=6.7$, corresponding to a parameter point that generates the observed DM relic abundance. For comparison, the corresponding evolution for $f_a=10$ and $10^3\,\mathrm{GeV}$  are also given. We always fix $m_\rho=2.5m_\pi$, $g_\rho=5$, $c_{\rho\rho}=\frac 12$ and $\mathsf{K}=-\frac{1}{4}$.}
\label{fig:evolution}
\end{figure}

{\bf Numerical results.\,}
With all the relevant interactions provided above, we now numerically solve the  dark  pion freeze-out, using the Boltzmann equations that describe the evolution of the comoving number/energy densities of the strong QCD/meson  sector:
\begin{align}
\frac{dn_{\pi}}{dt} + 3 H n_{\pi} =& -  \braket{\sigma v^2}_{3\pi \rightarrow 2\pi}  n_{\pi}^2\left(n_{\pi}- n_{\pi}^\text{eq}\right)\,,
\label{eq:dnpi_dt0} \\
	\frac{d\rho_s}{dt} + 3 H (\rho_s + p_s)  & = - \mathcal{Q}(n_\pi,T_\pi,T_a)\,,
\label{eq:drhos_dt0}
\end{align}
where $\mathcal{Q}(n_\pi,T_\pi,T_a)$ denotes the energy exchange between the mesons and axion thermal bath via $2\pi \leftrightarrow 2\pi + a $, depending on the temperatures of the meson and axion sectors. The Hubble rate $H$ is calculated from the total energy density including the SM bath and the meson and axion sectors. Due to the large coupling $g_\rho$, dark pions and $\rho$-mesons are always thermalized in the meson sector.  To simplify the computation, we only take the non-relativistic limit of each interaction rate, and further assume that the axion sector is self-thermalized~\footnote{While this may be a strong assumption, it should not affect the results qualitatively, due to $\rho_s \ll \rho_a$  during a non-relativistic freeze-out.}.  Exact expressions of these  relevant quantities and a full set of the Boltzmann equations are provided in the supplementary material.

We set that initially the pions and axions are coupled in a dark sector, being colder than the SM bath, with $T_\pi=T_a=0.7\, T_\gamma$. This is to avoid early Universe bounds on extra radiation~\cite{Dicus:1982bz,Schramm:1997vs,Planck:2018vyg}. During freeze-out, the $3\pi\to2\pi$ self-annihilation process  ejects kinetic energy to the pion bath, potentially increasing $T_\pi$ relative to the axion temperature $T_a$. This energy can be efficiently transferred to the radiation bath if kinetic equilibrium between pions and axions is maintained through processes such as $2\pi \leftrightarrow 2\pi+a$, ensuring $T_\pi\simeq T_a$ during freeze-out. Consequently,  maintaining  kinetic equilibrium until $m_\pi/T_s \gtrsim 15\text{--}20$ would be enough to sufficiently reduce the dark pion abundance, yielding a successful SIMP freeze-out.   

The numerical solution of  Eqs.~(\ref{eq:dnpi_dt0}-\ref{eq:drhos_dt0}) is illustrated in ~\autoref{fig:evolution} for the parameter set $m_\pi=100\,\mathrm{GeV}$, $m_\rho=2.5m_\pi$, and $m_\pi/f_\pi=6.7$, with  three benchmark values of $f_a$, $10\,\mathrm{GeV}$ (green lines), $10^2\,\mathrm{GeV}$ (blue) and $10^3\,\mathrm{GeV}$ (red). The top and bottom panels show the evolving temperature of each sector and the evolving dark pion abundance $Y_\pi = n_\pi/s$, respectively,  as a function of $x= m_\pi/T_\gamma$, where $s$ is the entropy density and  $T_\gamma$ is the photon temperature.   The horizontal dotted lines indicate the moment of kinetic decoupling,  $x_k$, in the upper panel and the moment of freeze-out, $x_f$, in the lower panel~\footnote{To characterize kinetic decoupling and DM freeze-out, we define $x_k$ as the point at which $T_\pi$ and $T_a$ differ by $1\%$, and $x_f$ as the point at which $Y_\pi$ and $Y_\pi^{\rm eq}$ differ by $1\%$.}.  For $f_a=10^3\,\mathrm{GeV}$, dark mesons kinetically decouple from the axion bath at $x_k\sim13$. After kinetic decoupling, self-annihilation process of $3\pi \to 2\pi$ reheats $T_\pi$ with respect to $T_a$, and in turn suppresses further SIMP freeze-out. Increasing axion-pion coupling thus makes kinetic equilibrium last longer, so the SIMP freeze-out  happens successfully with $T_\pi \simeq T_a$~\footnote{$T_\pi $ may become slightly larger than $ T_a$, if the kinetic energy ejection by $3\pi \to 2\pi$ dominates over energy dissipation processes $2\pi \leftrightarrow 2\pi  +a$, even if both rates are larger than the Hubble rate. }. This is the case for $f_a =10$ and $10^2$\,GeV, where the kinetic decoupling happens after freeze-out with $x_f \simeq 14$.  After kinetic decoupling and freeze-out, the temperature of non-relativistic dark pions deceases as $1/x^2$.

\begin{figure}[t]
    \centering
    \includegraphics[width=0.49\textwidth]{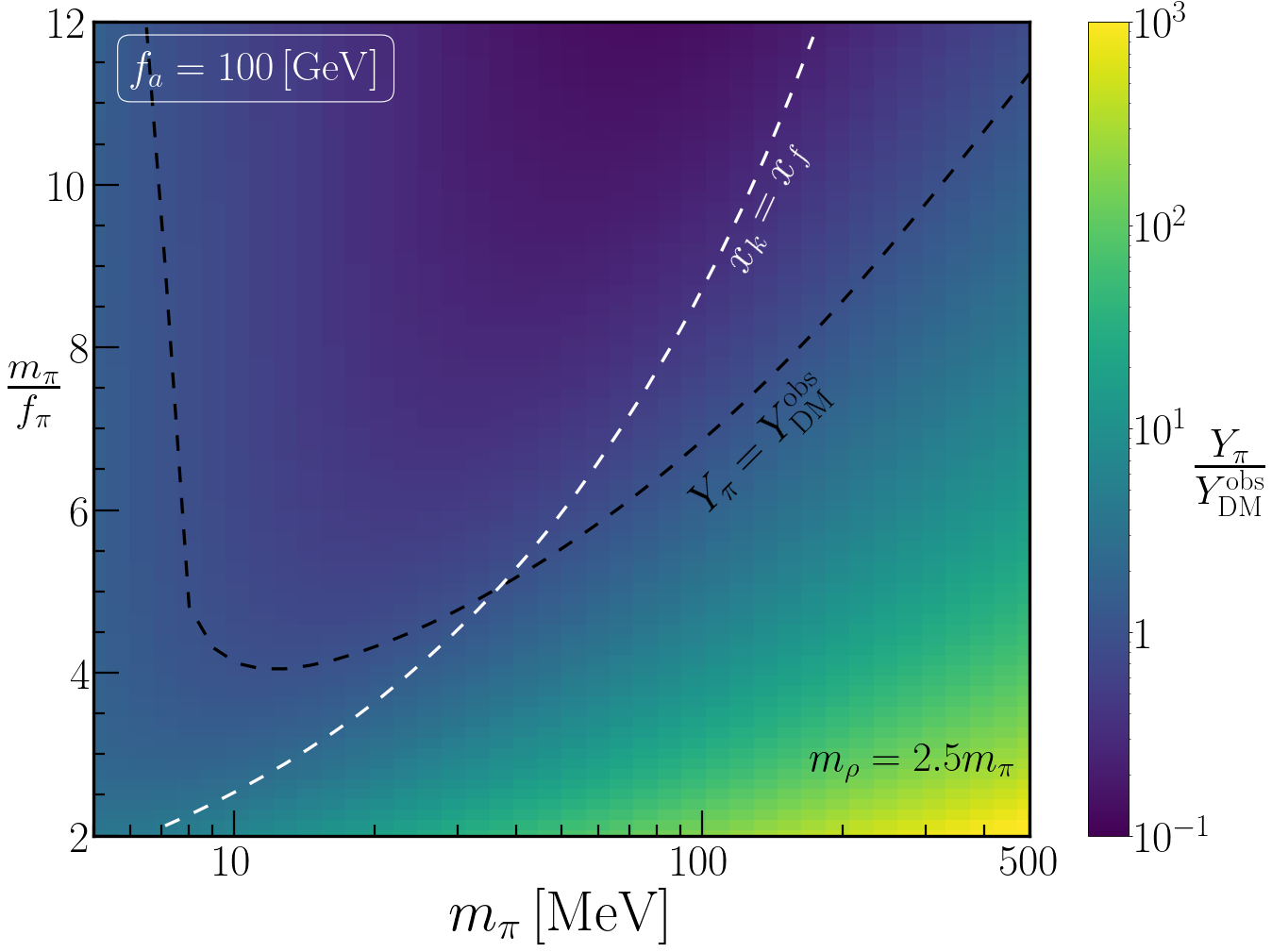}
    \caption{Values of $Y_\pi/Y_\pi^{\rm obs}$ for a  parameter scan in the ($m_\pi$,\,$m_\pi/f_\pi$) plane with $f_a=10^2\,\mathrm{GeV}\gg f_\pi$.  The black and white dashed lines denote, respectively, the parameter points generating the observed DM relic abundance and those for which kinetic decoupling occurs at freeze-out, $x_k=x_f$. We adopt  $m_\rho=2.5m_\pi$, $g_\rho=5$, $c_{\rho\rho}=\frac 12$ and $\mathsf{K}=-\frac{1}{4}$, same as~\autoref{fig:evolution}.}
    \label{fig:scanPara}
\end{figure}

In~\autoref{fig:scanPara}, we scan the ($m_\pi$,\,$m_\pi/f_\pi$) plane to calculate the normalized DM relic abundance $Y_\pi/Y_\pi^{\rm{obs}}$ for  $f_a=10^2\,\rm{GeV}$. The black dashed line denotes the parameter points for which the observed DM relic abundance is obtained, and the white dashed line corresponds to $x_k=x_f$. Below the white line, axions and pions remain in kinetic equilibrium during freeze-out, where efficient energy transfer to the axion bath allows significant kinetic energy dissipation, enabling freeze-out analogous to the standard SIMP mechanism~\cite{Hochberg:2014dra,Hochberg:2014kqa,Hansen:2015yaa, Braat:2023fhn}. In this case,   larger values of $m_\pi/f_\pi$ yield smaller final DM abundances.  Above the white line, kinetic decoupling happens earlier than DM freeze-out. As  $T_\pi$ approximately scales as $\ln T_\gamma$ after kinetic decoupling, increasing $m_\pi/f_\pi$ can barely enhance the freeze-out process. In contrast, now the final $Y_\pi$  mostly relies on $x_k$, and thus on the dimensionless coupling $m_\pi/f_a$. Deep in this parameter region, a smaller $m_\pi/f_a$, or equivalently a smaller $m_\pi$ here,  slows down the $3\pi \to 2\pi$ process at an earlier time, increasing  the final value of $Y_\pi$. As a result, a new parameter region opens up to yield the observed relic abundance.

The parameters that yield the observed DM abundance potentially also have   observable consequences. For instance, similar to Ref.~\cite{Farina:2016llk} the parameter region with an earlier kinetic decoupling for MeV-scale SIMP can provide warm DM, suppressing astrophysical small-scale structures. Our scenario can also exhibit potential gravitational waves from the confinement dark sector~\cite{Halverson:2020xpg, Reichert:2021cvs, Houtz:2025ogg, Morgante:2022zvc}, as well as from the breaking of $U(1)_{\rm{PQ}}$~\cite{DelleRose:2019pgi,Ghoshal:2020vud, VonHarling:2019rgb}. The axion itself has to be sub-MeV in our scenario, and thus contributes to extra radiation at big bang nucleosynthesis.

{\bf Conclusions.\,}
It was recognized soon after the first proposal of DM freeze-out via self-reducing processes that this scenario has an energy dissipation problem~\cite{deLaix:1995vi}. This is typically solved in the SIMP framework by introducing elastic scattering between DM and radiation~\cite{Hochberg:2014dra}. However, this solution suffers from its accompanied DM pair annihilation due to crossing symmetry, largely reducing its relevant parameter region.   
In this Letter, we propose internal Bremsstrahlung as a novel mechanism for maintaining kinetic equilibrium between SIMP particles and the radiation bath. We show that this mechanism does not introduce potentially dangerous DM pair annihilation processes\,\footnote{In contrast,  initial/final state radiation (ISR/FSR) processes typically induce DM pair annihilation. }, and can be naturally realized in a class of dark pion models with a radiative axion. 
Regarding the parameter region, this concrete  model contains two parameter regimes that yield the observed DM relic abundances for DM masses below GeV. One is the same as the conventional SIMP, where the final freeze-out happens before kinetic decoupling between dark pions and axions, with $x_k \gtrsim x_f$. For the reverse time ordering, $x_f \lesssim x_k$,  the DM abundance only reduces mildly after $x_k$, although the freeze-out may happen much later. Therefore, one can tune the time of kinetic decoupling to modify the final DM abundance. 

At the very end, it is worth emphasizing that this simplified dark pion model serves only as an example to illustrate our   mechanism for maintaining  kinetic equilibrium. Extra excited meson states do exist in such dark sectors and may affect the detailed evolution of the DM abundance. Nevertheless, extra states generally enhance relevant energy-transfer and/or DM annihilation rates, therefore making successful freeze-out even easier to achieve~\cite{Bernreuther:2023kcg, Chu:2024rrv, Garcia-Cely:2024ivo,Garcia-Cely:2025flv}. At the model-building level, our mechanism can be straightforwardly applied to other SIMP candidates, such as the dark vector DM model~\cite{Bernal:2015ova}. The original set-up there can be extended to couple the vector DM  particle directly to a heavy CP-even scalar, but not to a light CP-odd  pseudoscalar, in analogy to two-Higgs doublet DM models. Consequently, the scalar (pseudoscalar)  plays the role of the $\rho$ meson (axion) here. At the conceptual level, our study suggests a general structure, in which the DM candidate and the radiation bath are only coupled through predominantly off-shell heavier states, to achieve the desired kinetic equilibrium.  A similar structure, introducing a sub-leading DM component as the bridge instead, is considered in Ref.~\cite{Liu:2019knx} for a different purpose. These aspects, although deserving of further investigation, lie beyond the scope of this letter and are deferred to future work.

\section*{Acknowledgments}
The authors thank Jia Liu, Xiaolin Ma, and Josef Pradler for useful discussions. This work is funded by the National Natural Science Foundation of China (grant No. E4146602 and No. 12547104), and the Fundamental Research Funds for the Central Universities (grant No. E4EQ6605X2 and E5ER6601A2). Y.E.M. is also supported by the CAS-ANSO Scholarship for Young Talents. 
\end{sloppypar}

\bibliography{reference}

\appendix
\setcounter{secnumdepth}{2}

\onecolumngrid
\clearpage
\section{Model Building}
\label{sec:Model}
As an explicit example to illustrate our novel mechanism to achieve kinetic equilibrium  between the SIMP particles and radiation, we consider a dark QCD  model based on an $Sp(N_c)$ color gauge theory with $N_f$ Dirac fermions, $q_i$. This class of SIMP models was originally proposed in~\cite{Hochberg:2014kqa} and has since been extensively studied in the literature; see e.g. ~\cite{Hochberg:2015vrg,Braat:2023fhn,Kulkarni:2022bvh,Chu:2024rrv}. 

In the massless limit, the $2N_f$ Weyl  spinors from $N_f$ Dirac fermions, $q_i$, exhibits a global $SU(2N_f)$ flavor symmetry, due to the pseudo-real fundamental representation of the color group. 
To render the global $SU(4)$ flavor symmetry explicit, it is convenient to express the quark fields in the Weyl basis. The left- and right-handed components of the Dirac fermions can then be combined into a four-component spinor transforming in the fundamental representation of $Sp(4)$ using the Nambu–Gorkov formalism~\cite{Kulkarni:2022bvh, PhysRev.117.648, osti_7264935},
\begin{equation}
\Psi\equiv\begin{pmatrix}
u_L \\
 d_L\\
 \sigma_2 S u_R^\star\\
\sigma_2 S d_R^\star
\end{pmatrix}\,,
\end{equation}
where $S$ acts on the color group, with $S^2=-\mathbb{1}$ and $(\tau_a)^T=S \tau_a S$ for all generators $\tau_a$ in the $Sp(4)_c$ color group. The mass and kinetic term of the quarks can be expressed as~\cite{Kulkarni:2022bvh, Kogut:2000ek},
\begin{equation}
\mathscr{L}_{q}=i \Psi^\dagger \bar\sigma_\mu D^\mu \Psi+\left( -\frac12 \Psi^T \sigma_2 M_q  \Psi+\mathrm{h.c.} \right)\,,
\label{eq:quark_weyl}
\end{equation}
where $\bar{\sigma}=(1, -\vec{\sigma})$, with $\sigma_j$ are the Pauli matrices and $M_q(a)=i\sigma_2 \otimes m_q$. The kinetic term in~\autoref{eq:quark_weyl} exhibits an $U(4)$ symmetry, which is explicitly broken by non-zero quark masses. In analogy with QCD in the Standard Model (SM), the axial anomaly further reduces the symmetry, yielding an $SU(4)$ global flavor symmetry. 
After color confinement, this global symmetry is subsequently broken to $Sp(2N_f)$ by the chiral condensate,
\begin{equation}
\left\langle \bar q_i q_j \right\rangle= \mu^3 \Sigma_{0_{ij}}\,,
\end{equation}
where $\mu$ is a mass-dimension-one parameter, and $\Sigma_{0}=i \sigma_2 \otimes \mathbb{1}_{N_f}$ is a $2N_f \times 2N_f$ antisymmetric matrix, which is invariant under $Sp(2N_f)$ subgroup of $SU(2N_f)$. The chiral symmetry breaking gives rise to $N_\pi=2N_f^2-N_f-1$ pions, $\pi^a$,  which are  pseudo-goldstone bosons associated with the broken generators $T^a$ of the coset space $SU(2N_f)/Sp(2N_f)$, with $a=1,.., N_\pi$. 
In the approach of chiral effective theories, these pion fields can be written as fluctuations of the orientation of the chiral condensate,
\begin{equation}
\Sigma=e^{i \pi/f_\pi}\Sigma_0 e^{i \pi^T/f_\pi}\,,
\end{equation}
where $f_\pi$ is the pion decay constant and $\pi=\sum_a \pi_a T^a$, with $\Tr[{T^a T^b}]= \frac{1}{2} \delta ^{ab}$. These pions can be stable and thus serve as the DM candidate. 

Following the literature, this section summarizes the relevant Lagrangian terms before and after color confinement, once a  dark axion, or equivalently, an anomalous global $U(1)_\text{PQ}$, is present.  For concreteness, we take $N_c =4$ and $N_f =2$ and assume that in the dark sector there is only one gauge group, the dark color $Sp(4)_c$.

\subsection{The effective quark-level Lagrangian for axions}
We denote the two Dirac fermions as $u$ and $d$, and then introduce  to the minimal model a dark axion, which is supposed to (nearly) massless and evolves as radiation at early Universe. The general quark-axion effective Lagrangian above the confinement scale can be written as,
\begin{align}
 \mathscr{L}_{eff}=&\frac12 \partial_\mu a \partial^\mu a -\frac{m^2_{a,0}}{2}a^2+\bar{q}\left( i \slashed{D}-\textbf{m}_{q,0} \right)q+ \frac{\partial_\mu a}{f_a}\left(\mathbb{k}_{L,0} \bar{q}_L\gamma^\mu q_L+\mathbb{k}_{R,0} \bar{q}_R\gamma^\mu q_R  \right)\nonumber\\&+c_{\mathrm{gg}}\frac{\alpha}{4\pi}\frac{a}{f_a}G_{\mu \nu}\tilde{G}^{\mu\nu}+\frac{a}{f_a}\sum_{\mathbb{A}_{1,2}}c^0_{\mathbb{A}_{1}\mathbb{A}_{2}}F_{\mathbb{A}_{1}\mu\nu}\tilde{F}_{\mathbb{A}_{2}}^{\mu \nu}\,,
 \label{eq:quarkeff}
\end{align}
where the quark mass $\mathbf{m}_{q,0} = \mathrm{diag}(m_u, m_d)$, $\slashed{D}=\gamma^\mu D_\mu$, with $D_\mu$ being the covariant derivative, $G_{\mu \nu}$ denotes $Sp(4)_c$ color field strength. The field $\mathbb{A}$ denotes the gauge fields associated with gauge groups other than the color group, with $F_{\mathbb{A}\mu\nu}$ denoting their corresponding field strength and the dual $\tilde{F}_{\mu \nu}=\frac12 \epsilon^{\mu \nu \rho \sigma} F_{\rho \sigma}$. In our kinetic-equilibrium mechanism, we do not consider any gauge symmetry beyond the color group $Sp(4)_c$, but in the supplementary material retain $\mathbb{A}$ for a general discussion. 
When $m_u=m_d$, the quark Lagrangian exhibits a global $SU(4)$ flavor symmetry~\cite{Kulkarni:2022bvh, Kogut:2000ek}. To preserve this symmetry in the presence of the axion, we require that the axions have identical coupling to each quark flavor $\mathbb{k}_{L/R,0}=\rm{diag}(k_{L/R, 0},\, k_{L/R, 0})$, with $k_{L,0}=-k_{R,0}$. 

The axion-quark interaction can be embedded into the covariant derivative $D_\mu$ by incorporating axion field into a background fields $\mathbb{B}$, as
\begin{equation}
\mathscr{A} \equiv \mathbb{A}+\mathbb{B}= \mathbb{A}  + \left( \mathbb{K}\frac{da}{f_a}+ \mathbb{B}^V+\mathbb{B}^A\right) \,, \label{eq:bkg}
\end{equation}
where $\mathbb{K}=\mathrm{diag}\left\{ k_L(a), -k_R(a) \right\}$.
Additionally, following Ref.~\cite{Bai:2024lpq} we introduce $\mathbb{B}^{(A)V}$, the (axial-)vector meson fields, as background fields. Below the confinement scale, they serve as fictitious, non-dynamical gauge fields associated with the global flavor symmetry~\cite{Harvey:2007ca}. This allow us to track global anomalies and derive the counter-terms induced by the axion-dependent chiral rotation~\cite{Harvey:2007ca, Bai:2024lpq, Bai:2025fvl, Bai:2023bbg, Son:2004tq}. 
Adding background fields into gauge fields in~\autoref{eq:bkg}  breaks the underlying gauge invariance. To restore the gauge invariance we add counter-terms $\Gamma_c=\int d^4 x \mathscr{L}_c (\mathbb{A}_{L,R}, \mathbb{B}_{L,R})$~\cite{Harvey:2007ca} into~\autoref{eq:quarkeff}, in terms of 
\begin{equation}
\Gamma_c=-2 \mathscr{C}\int {\mathrm{Tr}\left[ \left( \mathbb{A}d \mathbb{A}+d \mathbb{A} \mathbb{A} \right) \mathbb{B}+\frac12 \mathbb{A}\left(  \mathbb{B}d \mathbb{B}+d \mathbb{B} \mathbb{B}\right)-\frac32i \mathbb{A}^3 \mathbb{B}-\frac34i \mathbb{A} \mathbb{B} \mathbb{A}  \mathbb{B} -\frac i2 \mathbb{A}  \mathbb{B}^3\right]} 
\end{equation}
with the coefficient  $\mathscr{C}=N_c/(48\pi^2)$.

To match the standard chiral Lagrangian, one can absorb the $aG\tilde{G}$ term by performing an axion-dependent phase rotation~\cite{Bauer:2021wjo}:
\begin{equation}
q(x) \rightarrow \mathrm{exp}\left[ -i\left( \mathbb{\delta}_{q,0}+{\kappa}_{q,0}\gamma_5 \right) c_{gg} \frac{a(x)}{f_a} \right]q(x)\,   \quad \text{with}\, \, \mathrm{Tr}(\kappa_{q,0})=1\,,
\label{eq:chiral_rot}  
\end{equation}
where $\mathbb{\delta}_{q,0}$ and $\kappa_{q,0}$ are diagonal matrices, with $\rm{Tr}[\kappa_{q,0}]=1$. Equivalently, the quark rotation can also be parametrized in the chiral basis with  ${\theta}_{L/R,0}=\mathbb{\delta}_{q,0} \mp  \kappa_{q,0}$. Preserving the explicit global $SU(4)$ symmetry constrains the allowed chiral rotation to ${\delta}_{q,0}=0$ and $\kappa_{q,0}=\kappa\,\mathbb{1}_2$. The quark rotation induces the shifts $\textbf{m}_{q,0} \to \textbf{m}_q(a)$,  $\textbf{k}_{L/R,0} \to \textbf{k}_{L/R}(a)$ and $c^0_{\mathscr{A}_{1}\mathscr{A}_{2}} \to c_{\mathscr{A}_{1}\mathscr{A}_{2}}$~\cite{Bauer:2021wjo},  and modifies \autoref{eq:quarkeff} to 
\begin{equation}
 \mathscr{L}_\text{eff}=\frac12 \partial_\mu a \partial^\mu a -\frac{m^2_{a,0}}{2}a^2+\bar{q}\left( i \slashed{D}-\textbf{m}_{q}(a) \right)q+\frac{a}{f_a}\sum_{\mathscr{A}_{1,2}}c_{\mathscr{A}_{1}\mathscr{A}_{2}}F_{\mathscr{A}_{1}\mu\nu}\tilde{F}_{\mathscr{A}_{2}}^{\mu \nu}+\mathscr{L}_c\,.
\end{equation}
These axion-dependent parameters, as already given by Ref.~\cite{Bai:2024lpq}, are 
\begin{align}
m_q(a)&=e^{ic_{gg}\theta_{L,0}a/f_a}m_{q,0}e^{-ic_{gg}\theta_{R,0}a/f_a}\,,\\
\textbf{k}_L(a)&=e^{ic_{gg}\theta_{L,0}a/f_a}\left[\textbf{k}_{L,0}+c_{gg} \theta_{L,0} \right]e^{-ic_{gg}\theta_{L,0}a/f_a}\,,\\
\textbf{k}_R(a)&=e^{ic_{gg}\theta_{R,0}a/f_a}\left[\textbf{k}_{R,0}+c_{gg} \theta_{R,0} \right]e^{-ic_{gg}\theta_{R,0}a/f_a}\,,\\
\frac{a}{f_a}\sum_{\mathscr{A}_{1,2}}c_{\mathscr{A}_{1}\mathscr{A}_{2}}F_{\mathscr{A}_{1}\mu\nu}\tilde{F}_{\mathscr{A}_{2}}^{\mu \nu}&=\frac{a}{f_a}\sum_{\mathscr{A}_{1,2}}c^0_{\mathscr{A}_{1}\mathscr{A}_{2}}F_{\mathscr{A}_{1}\mu\nu}\tilde{F}_{\mathscr{A}_{2}}^{\mu \nu}-\delta\left[ \mathscr{L}_{\mathrm{WZW}}+\mathscr{L}_{\rm{eff}} \right]\left( c_{gg} \theta_{L,0},c_{gg} \theta_{R,0} \right)\,. \label{eq:cff_ax}
\end{align}
If $k_{L/R,0}$ are diagonal matrices, the couplings $k_{L/R}$ do not acquire any dependence on the axion field and are given by $k_{L/R}=k_{L/R,0}+c_{gg} \theta_{L/R,0}$. To preserve gauge invariance under the axion rotation in the presence of background fields, we introduce the counter-terms $\delta\left[ \mathscr{L}_{\mathrm{WZW}}+\mathscr{L}_{\rm{eff}}  \right]$ in \autoref{eq:cff_ax}, such that
$\delta \Gamma_c=-\delta\left[ \mathscr{L}_{\mathrm{WZW}}+\mathscr{L}_{\rm{eff}}  \right]$, given by~\cite{Harvey:2007ca}:
\begin{align}
\delta\left[ \mathscr{L}_{\mathrm{WZW}}+\mathscr{L}_{\rm{eff}}  \right]=-2 \mathscr{C}\frac {a}{f_a}\int \rm{Tr}\bigg[ \theta\bigg(&3(d\mathbb{A}-i\mathbb{A}^2)^2+3(d\mathbb{A}-i\mathbb{A}^2)D\mathbb{B}+(D\mathbb{B})^2 -\frac {i}{2}D(\mathbb{B}^3) +i\mathbb{B}(d\mathbb{A}-i\mathbb{A}^2)\mathbb{B}-i(d\mathbb{A}-i\mathbb{A}^2)\mathbb{B}^2\bigg) \bigg]\,,
\end{align}
where ${\theta}=\rm{diag}(\theta_L,-\theta_R)$,  $D\mathbb{B}=d\mathbb{B}-i\mathbb{A}\mathbb{B}-i\mathbb{B}\mathbb{A}$,  and $d\mathbb{A}-i\mathbb{A}^2$ is the covariant field strength of gauge fields. 

\subsection{Chiral symmetry breaking}
As already mentioned above, the global $SU(4)$ flavor symmetry is spontaneously broken to $Sp(4)$ through the formation of a chiral condensate $\left\langle\bar{u}u+\bar{d}d\right\rangle= \mu^3 \Sigma_{0_{ij}}$, resulting in five pions (pseudo-Goldstone bosons), which can be represented as:
\begin{equation}
\pi=\sum_{a=1}^5 \pi_a T_a=\frac{1}{2\sqrt2}\begin{pmatrix}
\pi_3 &\pi_1-i\pi_2  &0  & \pi_5-i\pi_4  \\
\pi_1+i\pi_2 & -\pi_3  &-\pi_5+i\pi_4  & 0  \\
0 & -\pi_5-i\pi_4 & \pi_3  & \pi_1+i\pi_2 \\
\pi_5+i\pi_4 & 0  & \pi_1-i\pi_2 & -\pi_3
\end{pmatrix}\,.
\label{eq:pi}
\end{equation}
Analogous to the SM QCD theory, the $SU(4)$ symmetry results in a ten-plet of vector mesons $\rho$  (an even heavier quintuplet of axial-vector mesons  $\omega$), associated with the unbroken (broken) generators of $SU(4)$~\cite{Kulkarni:2022bvh}. These  mesons can be parameterized as 
\begin{align}
\rho^\mu&=\sum_{n=6}^{15} \rho_n T_n=\frac{1}{2\sqrt2}\begin{pmatrix}
\rho_{14}+\rho_{15} &\rho_{13}-i\rho_{8}  &\sqrt{2}\rho_{10}-i\left( \rho_{6}+\rho_{9} \right)  & \rho_{11}-i\rho_{7}  \\
\rho_{13}+i\rho_{8} & \rho_{15}-\rho_{14}  &\rho_{11}-i\rho_{7}  & \sqrt{2}\rho_{12}-i\left( \rho_{6}-\rho_{9} \right)  \\
\sqrt{2}\rho_{10}+i\left( \rho_{6}+\rho_{9} \right) & \rho_{11}+i\rho_{7}  & -\left( \rho_{14}+\rho_{15} \right) & -\left( \rho_{13}+i\rho_8 \right) \\
\rho_{11}+i\rho_{7} & \sqrt{2}\rho_{12}-i\left( \rho_{6}-\rho_{9} \right)  & -\left( \rho_{13}-i\rho_8 \right) & -\left( \rho_{15}-\rho_{14} \right)
\end{pmatrix}^\mu\,,
\label{eq:rho}\\
\omega^\mu&=\sum_{a=1}^5 \omega_a T_a=\frac{1}{2\sqrt2}\begin{pmatrix}
\omega_3 &\omega_1-i\omega_2  &0  & \omega_5-i\omega_4  \\
\omega_1+i\omega_2 & -\omega_3  &-\omega_5+i\omega_4  & 0  \\
0 & -\omega_5-i\omega_4 & \omega_3  & \omega_1+i\omega_2 \\
\omega_5+i\omega_4 & 0  & \omega_1-i\omega_2 & -\omega_3
\end{pmatrix}^\mu\,.
\end{align}
In this work, we impose flavor universality, so pions and vector mesons are degenerate in mass, with their  masses denoted as $m_\pi$ and $m_\rho$, respectively. The axial-vector mesons are taken to be heavier than the vector ones, and therefore do not play a significant role in the SIMP mechanism. In practice, their presence only enhance the energy transfer between the meson and axion sectors.

\subsection{The Chiral Lagrangian and  Wess–Zumino–Witten Interactions in presence of an axion}

Below the confinement scale, the dynamics of mesons and axions can be described by the effective chiral Lagrangian together with the WZW term. The chiral Lagrangian is provided by:
\begin{equation}
\mathscr{L}_{\chi\mathrm{PT}}=\frac{f^2_\pi}{4}\mathrm{Tr}\left[ (D^\mu \Sigma )(D_\mu \Sigma)^\dagger \right]-\frac{f^2_\pi}{2}B_0\mathrm{Tr}\left[ M_q(a)\Sigma+\mathrm{h.c.} \right]+\frac12 \left( \partial_\mu a \right)\left( \partial^\mu a \right)-\frac{m^2_{a,0}}{2}a^2+ \frac{a}{f_a} \sum_{\mathscr{A}_{1,2}}c_{\mathscr{A}_1 \mathscr{A}_2}F_{\mathscr{A}_1 \mu \nu}\tilde{F}_{\mathscr{A}_2}^{\mu \nu}\,,
\label{eq:chiPT}
\end{equation}
where $B_0\approx m^2_\pi/(m_u+m_d)$. The covariant derivative $D_\mu$ now has the form of 
\begin{equation}
D_\mu \Sigma \approx \partial_\mu \Sigma+ig_\rho\left( \rho_\mu \Sigma+\Sigma\rho_\mu^T \right)+i\frac{\partial_\mu a}{f_a}\left(\mathbb{K} \Sigma+\Sigma \mathbb{K}^T  \right)\,,
\end{equation}
where $\mathbb{K}=\mathrm{diag}\left\{ \mathbb{k}_L, -\mathbb{k}_R \right\}\equiv \mathsf{K}\cdot \mathbb{1}_4$. Here, we have neglected the  contributions of other gauge bosons and heavier axial-vector mesons, which are above the energy scales of interest.
It is well known now that the Wess-Zumino-Witten (WZW) term~\cite{Wess:1971yu, Witten:1983tw} is nonvanishing for $N_f \geq2 $,  inducing a five-point interaction between pions as 
\begin{equation}
\Gamma_{\mathrm{WZW}}^{0}=-\frac{iN_c}{240\pi^2}\int \left( \Sigma^\dagger d \Sigma \right)^5\,.
\end{equation}
This term enables $3\pi \to 2\pi$ self-annihilation process, which regulates the dark pion abundance during freeze-out, referred to as the ``SIMP mechanism''. 
Now the full WZW term has to include the original WZW term and the counter-terms, which are required to preserve gauge invariance. In terms of differential one-forms, the full WZW interaction terms are given by~\cite{Molinaro:2017mwb, Duan:2000dy}:
\begin{align}
\Gamma_{\mathrm{WZW}}=&\Gamma^0_{\mathrm{WZW}}+5\mathscr{C}\int \mathrm{Tr}\left\{ 2i\mathscr{A}\alpha^3-2\left( \mathscr{A}d\mathscr{A}+d\mathscr{A}\mathscr{A} \right)\alpha -\left[ d\mathscr{A}d\Sigma \mathscr{A}^T \Sigma^\dagger-d\mathscr{A}^Td\Sigma^\dagger \mathscr{A} \Sigma \right]\right. \nonumber\\
&-\left[ \Sigma \mathscr{A}^T \Sigma^\dagger \left( \mathscr{A}\alpha^2+\alpha^2 \mathscr{A} \right) \right]+\left( \mathscr{A}\alpha \right)^2+2i\left( \mathscr{A}^3 \alpha \right)+2i(d\mathscr{A}A+Ad\mathscr{A}) \Sigma \mathscr{A}^T \Sigma^\dagger \nonumber\\
&-2i(\mathscr{A}\alpha\mathscr{A}\Sigma \mathscr{A}^T\Sigma^\dagger)
\left. +2\left( \mathscr{A}^3 \Sigma \mathscr{A}^T \Sigma^\dagger \right)+\frac12\left( \mathscr{A} \Sigma \mathscr{A}^T \Sigma^\dagger \right)^2 \right\}\,,
\label{eq:fullWZW}
\end{align}
where $\mathscr{C}=\frac{iN_c}{240\pi^2}$, $\alpha= D\Sigma \Sigma^\dagger$, and $\mathscr{A}=\mathbb{A}+\mathbb{B}$, including  gauge fields beyond the color group and background fields from (axial-)vector mesons and first derivative of the axion.

Now we calculate explicitly the relevant interaction vertices, $\pi \pi \rho$ and $\rho \rho a$,  which play the crucial role in our kinetic equilibrium mechanism. Same as the SM chiral effective theory, the former one, $\pi \pi \rho$, arises from the chiral Lagrangian in~\autoref{eq:chiPT} at the first order 
\begin{align}
\frac{f^2_\pi}{4}\left( D_\mu \Sigma \right)\left( D^\mu \Sigma \right)^\dagger&\supset\frac{f^2_\pi}{4}\left(\frac{2i}{f_\pi}\left[ \partial_\mu \pi \Sigma_0+i g_\rho\left( \rho_\mu \pi \Sigma_0+\pi \Sigma_0 \rho_\mu^T \right) \right] \right)\left( D^\mu \Sigma \right)^\dagger \nonumber\\
&=\frac{f^2_\pi}{4}\left(\frac{2i}{f_\pi}\left[ \partial_\mu \pi \Sigma_0+i g_\rho\left( \rho_\mu \pi \Sigma_0-\pi \rho_\mu J \right) \right] \right)\left( D^\mu \Sigma \right)^\dagger \nonumber\\
&\supset 2i g_\rho \mathrm{Tr}\left[ \partial_\mu \pi\left( \rho^\mu\pi-\pi \rho^\mu \right) \right]\,,
\label{eq:pipi_rho}
\end{align}
where we have used $\rho \Sigma_0+ \Sigma_0 \rho^T=0$,  as vector meson $\rho$ is associated with the generators of the unbroken $Sp(4)$ symmetry. This relation will also be used below. 

Regarding the $\rho \rho a$ interaction vertex, the chiral Lagrangian in~\autoref{eq:chiPT}  yields 
\begin{align}
 \frac{a}{f_a} \sum_{i,j} c_{\mathcal{A}^i \mathcal{A}^j} F_{\mu \nu}^i \tilde{F}^{j, \mu \nu} &\supset \frac{a}{f_a} c_{\rho \rho} F_{\rho, \mu \nu} \tilde{F}_\rho^{\mu \nu} \nonumber\\
&\supset\frac{a}{2f_a} c_{\rho \rho} g_\rho^2 \epsilon^{\mu \nu \lambda \sigma} \text{Tr}  \left[(\partial_\mu \rho_\nu - \partial_\nu \rho_\mu) (\partial_\lambda \rho_\sigma - \partial_\sigma \rho_\lambda) \right] \nonumber\\
&=\frac{2a}{f_a} c_{\rho \rho} g^2 \epsilon^{\mu \nu \lambda \sigma} \text{Tr} \left[ \partial_\mu \rho_\nu \, \partial_\lambda \rho_\sigma \right]=\frac{a}{f_a} c_{\rho \rho} g_\rho^2\epsilon^{\mu \nu \lambda \sigma} \sum_i \partial_\mu \rho_{i\nu} \, \partial_\lambda \rho_{i\sigma}\,,
\label{eq:chiral_rhorhoa}
\end{align}
where $i$ denotes $\rho$ meson flavor index.
In addition, the full WZW term in~\autoref{eq:fullWZW} also gives rise to this interaction:
\begin{align}
10i\mathscr{C} \int_{M^4} \mathrm{Tr}&\left[ \left( d\mathscr{A}\mathscr{A}  +\mathscr{A}d\mathscr{A} \right) \Sigma \mathscr{A}^T \Sigma^\dagger \right]  \nonumber\\
&\supset 10i\mathscr{C} g_\rho^2  \int_{M^4} \mathrm{Tr}\left[\left\{ \left( \partial \rho \mathbb{K} \frac{\partial a}{f_a}+\mathbb{K} \frac{\partial a}{f_a}\partial \rho \right) \Sigma_0 \rho^T \Sigma_0^\dagger \right\} +\left\{ \left( \partial \rho \rho+\rho\partial \rho \right) \Sigma_0  \mathbb{K} \frac{\partial a}{f_a} \Sigma_0^\dagger \right\} \right] \nonumber\\
&=20i\mathscr{C}g_\rho^2 \mathsf{K} \int_{M^4} d^4x\frac{a}{f_a}   \partial_\mu \rho_\nu^A \partial_\lambda \rho_\sigma^A \epsilon^{\mu \nu \lambda \sigma} \,. \label{eq:WZW_rhorhoa}
\end{align}

One might anticipate that a $\pi\pi\rho a$ vertex, which could be  relevant for our kinetic-equilibrium mechanism, arises from the full WZW term in~\autoref{eq:fullWZW}. However, their total contribution vanishes, after summarizing the following three relevant terms: 
\begin{align}
5\mathscr{C} \int_{M^4} \mathrm{Tr}\left[ \left( \mathscr{A \alpha} \right)^2 \right] &\supset -10\mathscr{C}g_\rho \left(\frac{2}{f_\pi}  \right)^2 \int_{M^4} \mathrm{Tr}\left[ {\mathbb{K}}\frac{\partial a}{f_a}\, \partial \pi\, \rho \, \partial\pi \right] \nonumber \\
&=10\mathscr{C}g_\rho\left(\frac{4}{f_\pi^2}  \right) \mathsf{K}\frac{a}{f_a}\int_{M^4} \mathrm{Tr}\left[  \partial\pi\, \partial \pi\, \partial\rho  \right] \,, \label{eq:wzwppra1}
\end{align}
\begin{align}
-5\mathscr{C}\int_{M^4}\mathrm{Tr}\left[ \Sigma A ^T \Sigma^\dagger\left( \mathscr{A}\alpha^2+\alpha^2\mathscr{A} \right) \right]&\supset5\mathscr{C}g_\rho\left( \frac{4}{f_\pi^2} \right)\int_{M^4}\mathrm{Tr}{\left[ -\rho\left( {\mathbb{K}} \frac{\partial a}{f_a}\partial \pi \partial\pi+\partial \pi \partial\pi{\mathbb{K}}\frac{\partial a}{f_a} \right)+J{\mathbb{K}} \Sigma_0^\dagger \frac{\partial a}{f_a}\left( \rho\partial \pi \partial\pi+\partial \pi \partial\pi \rho \right) \right]} \nonumber\\
&=-20\mathscr{C}g_\rho\left( \frac{4}{f_\pi^2} \right)\mathsf{K}\frac{a}{f_a}\int_{M^4}\mathrm{Tr}[\partial \pi\, \partial \pi\, \partial \rho]\,,
\end{align}
and 
\begin{align}
-5\mathscr{C}\int_{M^4}\mathrm{Tr}\left[ d\mathscr{A} d \Sigma A^T \Sigma^\dagger-d\mathscr{A}^T d \Sigma^\dagger A \Sigma \right]&\supset -5\mathscr{C}g_\rho\left(\frac{4}{f_\pi^2}  \right) \int_{M^4}\mathrm{Tr}{\left[ \partial \rho \partial \pi \Sigma_0{\mathbb{K}} \frac{\partial a}{f_a}\Sigma_0^\dagger \pi-\partial\rho^T \Sigma_0^\dagger \partial \pi{\mathbb{K}} \frac{\partial a}{f_a} \pi \Sigma_0 \right]} \nonumber\\
&=10\mathscr{C}g_\rho\left(\frac{4}{f_\pi^2}  \right) \mathsf{K}\frac{a}{f_a}\int_{M^4}\mathrm{Tr}\left[\partial\pi \,\partial \pi\,\partial \rho       \right]\,.
\end{align}
We have used the relation $\rho^T=-\Sigma_0^\dagger \rho \Sigma_0$ to obtain the final expressions.

\subsection{Vanishing direct couplings between axion and pions}

In contrast to the $SU(2)_L \times SU(2)_R$ model studied in Ref.~\cite{Bai:2024lpq}, where axion–pion mixing arises through the neutral pion, such mixing is absent in this framework. While this has been discussed in the main text based on the fact that  the five pions follow the adjoint antisymmetric representation of $Sp(4)$, here we demonstrate it by explicit calculations. 

Expanding the kinetic term in the Lagrangian of  \autoref{eq:chiPT}, we obtain the axion-pion kinetic term:
\begin{align}
\frac{f^2_\pi}{4}\mathrm{Tr}\left[ (D^\mu \Sigma )(D_\mu \Sigma)^\dagger \right] &\supset {f^2_\pi}\mathrm{Tr} \left[ \frac{\partial_\mu a}{f_a}\left( \mathbb{K}+\Sigma_0 \mathbb{K} \Sigma_0^\dagger \right) \frac {\partial^\mu \pi}{f_\pi} \right] \nonumber\\
&=2\frac{f_\pi}{f_a} \mathsf{K}\frac {\partial_\mu a}{f_a}\rm{Tr}[\partial^\mu \pi] =0\,, \label{kin_mixing}
\end{align}
which vanishes because $\rm{Tr}[T^a]=0$.
The mass term in the Lagrangian of  \autoref{eq:chiPT} gives rise to the following axion–pion mass mixing term:
\begin{align}
-\frac{f^2_\pi}{2}B_0\mathrm{Tr}\left[ M_q(a)\Sigma+\mathrm{h.c.} \right]\supset 2f_\pi \rm{Tr}\left[ \begin{pmatrix}
(\theta_L-\theta_R)c_{gg}\frac{a}{f_a} &0  \\
 0& (\theta_L-\theta_R)c_{gg}\frac{a}{f_a}
\end{pmatrix}\pi \right]=0\,, 
\end{align}
due to the same reason. Furthermore, the $\mathcal{O}(a\pi^3)$ interaction terms arising from both the kinetic and mass contributions vanish due to the fact $\rm{Tr[\pi^3]} =0$.  One may  anticipate an $\mathcal{O}(a \pi^3)$ term arising from the second term in~\autoref{eq:fullWZW}, which could potentially jeopardize the SIMP mechanism. However, this term reduces to a total derivative due to the antisymmetric nature of the wedge product,
\begin{align}
10i\mathscr{C} \int_{M^4}  \mathrm{Tr}\left[ \mathscr{A} \alpha^3 \right] &\supset10i\mathscr{C} \int_{M^4}d^4x \mathrm{Tr}\left[ \mathbb{K}\frac{\partial_\mu a}{f_a}  \partial_\nu \pi \, \partial_\rho \pi \, \partial_\sigma \pi \right]\epsilon_{\mu \nu \rho \sigma} \nonumber\\
&=-10i\mathscr{C} \int_{M^4} \mathrm{Tr}\left[ \mathbb{K}\partial_\mu\left( \frac{ a}{f_a}  \partial_\nu \pi \, \partial_\rho \pi \, \partial_\sigma \pi \right) \right]\epsilon_{\mu \nu \rho \sigma}\,.
\end{align}
The same argument can be applied to the $\rho\pi a$ interaction arising from the term $\mathrm{Tr}\left[ \left( d\mathscr{A} \mathscr{A}+\mathscr{A} d\mathscr{A} \right) \alpha \right]$.

The terms $a\pi^2$ and $a\pi^4$ violate parity and therefore cannot arise from the chiral Lagrangian. For instance, one might expect  $a\pi^2$ term to arise from the kinetic term in~\autoref{eq:chiPT}:
\begin{align}
\frac{f^2_\pi}{4}\mathrm{Tr}\left[ (D^\mu \Sigma )(D_\mu \Sigma)^\dagger \right] 
\supset&\frac{f^2_\pi}{4}{\mathrm{Tr}} \bigg[\left( i \frac{2\partial_\mu\pi}{f_\pi}\Sigma_0 \right)\left( -\frac{\partial^\mu a} {f_a}\Sigma_0 ^\dagger\frac {2\pi}{f_\pi}\mathbb{K}-\frac {\partial^\mu a} {f_a}\mathbb{K}\Sigma_0 ^\dagger\frac{2\pi}{f_\pi} \right)+\rm{h.c.}\bigg]\,, \nonumber\\
&={\mathrm{Tr}}\bigg[-i\frac {\partial^\mu a} {f_a}\left( \partial_\mu\pi \pi \mathbb{K}+  \pi \partial_\mu\pi \Sigma_0 \mathbb{K} \Sigma_0^\dagger\right) +\rm{h.c.}\bigg]\,, \nonumber\\
&=2{\mathrm{Tr}}\bigg[-i\frac{\partial^\mu a} {f_a}\left( \partial_\mu\pi \pi \mathbb{K}\right) +\rm{h.c.}\bigg]=0\,. \label{eq:kinpi2a}
\end{align}
It's important to emphasize that the term in~\autoref{eq:kinpi2a} vanishes only when the axion is a singlet under $Sp(4)$ symmetry.  Additionally, since parity is violated by the WZW term, one might expect an $a\pi^4$ interaction to arise from the $\mathscr{A}\alpha^3$ term in~\autoref{eq:fullWZW}. This contribution, however, vanishes due to the anti-symmetry of differential one-forms.  

\section{Boltzmann Equations}
\label{sec:Bolt}
In this section, we discuss the relevant Boltzmann equations governing the evolution of the number density and temperature of the dark sector in our analysis. During the time scales relevant for DM freeze-out, dark sector pions $\pi$ and mesons $\rho$ are non-relativistic, whereas the axion  is relativistic and effectively acts as radiation. For strong dynamics, the meson (inverse-)decay process $\rho \leftrightarrow 2\pi$ remains in thermal equilibrium throughout the freeze-out process, thereby maintaining both kinetic and chemical equilibrium between the pions $\pi$ and vector mesons $\rho$, which consequently share a common temperature $T_\pi$. All pions $\pi_i$ (vector mesons $\rho_i$) belong to the same $SU(4)$ multiplet and thus possess identical interactions; we therefore assume identical number densities for all pion (vector meson) species and define the total pion and vector meson number densities as,
\begin{equation}
n_{\pi} \equiv N_\pi n_{\pi_i}, \qquad
n_{\rho} \equiv N_\rho n_{\rho_i}\,,
\end{equation}
 where $N_\pi =5$ and $N_\rho =10$. The combined energy density and pressure of the meson sector are given by
\begin{equation}
\rho_s = \rho_\pi +\rho_\rho \approx \rho_\pi \approx n_\pi \left(m_\pi+ \frac{3}{2} T_\pi\right) ,  \quad \quad
P_s \approx P_\pi \approx  n_\pi T_\pi\,.
\end{equation}
Since we are interested in parameter space with $m_{\rho}>2 m_\pi$,  $n_\rho$ is exponentially suppressed compared to that of $n_\pi$ during the freeze-out process, and thus we neglect $n_\rho$.
As the axion behaves as a relativistic species, its energy density and pressure are provided by
\begin{equation}
\rho_a = \frac{\pi^2}{30} T_a^4\,\quad \quad \text{~and~} \quad  \quad
P_a =\frac{\pi^2}{90} T_a^4\,,
\end{equation}
where $T_a$ denotes axion temperature.  That is, since we take the $\rho$ mesons to be much more massive than the pions, their contribution can be safely ignored, thus we end up with a system with only pions and the axion.

In our mechanism, the DM relic abundance is governed by the $3\pi \to 2\pi$ self-annihilation, with the released kinetic energy released during this process transferred to axions through inelastic scattering $2\pi \leftrightarrow 2 \pi \,+\,a$, mediated by $\rho$ mesons.  Such internal Bremsstrahlung processes ensure a quick cooling of dark pions without the need for a SM portal. The Boltzmann equations governing the evolution of $n_\pi$ is given by,
\begin{equation}
\frac{dn_{\pi_i}}{dt} + 3 H n_{\pi_i} = - \sum_{j,k,l,m} \braket{\sigma v^2}_{{\pi_i}{\pi_j}{\pi_k} \rightarrow {\pi_l}{\pi_m}}  \left(n_{\pi_i}n_{\pi_j}n_{\pi_k}-n_{\pi_i}^{\rm{eq}}n_{\pi_j}^{\rm{eq}}n_{\pi_k}^{\rm{eq}} \frac{n_{\pi_l}n_{\pi_m}}{n_{\pi_l}^{\rm{eq}}n_{\pi_m}^{\rm{eq}}}\right)\,,
\label{eq:Bolt_npi}
\end{equation}
where $n_i^{\rm {eq}}$ denotes the Maxwell–Boltzmann equilibrium number density~\cite{Kolb:1990vq} 
\begin{align}
    n_i^{\rm {eq}}(g_i,m_i,T_i)=g_i \left(\frac{m_i T_i}{2 \pi}\right)^{\frac{3}{2}} \exp \left(-\frac{m_i}{T_i}\right)\,,
\end{align}
with $g_\pi=1$ for each dark pion flavor. We adopt the thermally-averaged annihilation cross section from Ref.~\cite{Kamada:2022zwb} as 
\begin{align}
\braket{\sigma v^2}_{3 \pi \rightarrow 2 \pi} = \frac{25 \sqrt{5} N_c^2 m_\pi^3}{512 \pi^5 f_\pi^{10} N_\pi^3} T_\pi^2\,,
\end{align}
with $N_c$ being the rank of the Symplectic group $Sp(N_c)$, where we set $N_c =4$.

Using energy conservation, the evolution equations for the energy densities of the strong sector ($\pi+\rho$) and the axions are given by 
\begin{align}
	\frac{d\rho_a}{dt} + 3 H (\rho_a + p_a) =& \mathcal{Q}(n_\pi,T_\pi,T_a)\,, 
\label{eq:drhoa_dt} \\
	\frac{d\rho_s}{dt} + 3 H (\rho_s + p_s) =&  - \mathcal{Q}(n_\pi,T_\pi,T_a) \,,
\label{eq:drhos_dt}
\end{align}
where $\mathcal{Q}$ denotes the energy exchange between the pion and axion thermal baths, acting to equalize their temperatures, $H= \sqrt{\frac{8 \pi G}{3}(\rho_{\mathrm{SM}} + \rho_{s}+\rho_{a})}$ is the Hubble parameter, with the SM energy density given by $\rho_{SM}(T_\gamma) = \frac{\pi^2}{30} g^E_\star(T_\gamma)  T^4$, with $T_\gamma$ being the SM temperature and $g^E_\star$ being the SM effective energy degrees of freedom. The energy transfer function $\mathcal{Q}$ can be further parameterized as
\begin{align}
\mathcal{Q}(n_\pi,T_\pi,T_a) \equiv \mathcal{T}\left(T_\pi,T_a\right) n_\pi^2\,,
\end{align}
in which the full expression of the function $\mathcal{T}$ will be provided later  in~\autoref{app:cross}.  

The time evolution of temperatures $T_\pi$ and $T_a$ can be obtained from \cref{eq:drhoa_dt,eq:drhos_dt} using the relation: 
\begin{equation}
\frac{d \rho (t, T_i)}{dt} = \frac{\partial \rho}{\partial n}  \frac{d n}{d t} + \frac{\partial \rho}{\partial T_i} \frac{d T_i}{d t}\,.
\end{equation}
Consequently, the equations governing the temperature evolution are given by 
\begin{align}
	 \frac{dT_\pi}{dt}  =& -2 H T_\pi + \frac{2}{3n_\pi}\Bigg[\left(m_\pi+\frac{3}{2}T_\pi\right) \braket{\sigma v^2}_{3 \pi \rightarrow 2 \pi} n_\pi^2 \left(n_\pi-n_{\pi,eq}\right)+ \mathcal{T} \left(T_\pi,T_a\right) \, n_\pi^2 \Bigg], \label{eq:dTs_dT}\\
	\frac{dT_a}{dt} =& - H T_a -\frac{15}{2 \pi^2 T_a^3} \mathcal{T} \left(T_\pi,T_a\right) \, n_\pi^2. \label{eq:dTa_dT}
\end{align}
For the numerical analysis, it is convenient to work with the comoving number density, $Y_X = n_X/s$, where $s$ denotes the entropy density of the SM particles, together with the dimensionless variable $x = m_\pi/T_\gamma $. Approximately, the system of equations in  \cref{eq:Bolt_npi,eq:dTs_dT,eq:dTa_dT} then simplifies to 
\begin{align}
    \frac{d\,Y_{\pi}(x)}{dx}=&-\frac{1}{H x} \braket{\sigma v^2}_{3 \pi \rightarrow 2 \pi} s^2 Y_\pi^2 \left(Y_\pi-Y_{\pi,eq}\right),\label{eq:Bolt_npi-final} \\
    \frac{d\,T_\pi(x)}{dx}  =&-\frac{2 T_\pi}{x}+\frac{2}{3 H x }\Bigg[ (m_\pi+\frac{3}{2}T_\pi) \braket{\sigma v^2}_{3 \pi \rightarrow 2 \pi} s^2 Y_\pi \left(Y_\pi-Y_{\pi,eq}\right)+ \mathcal{T}\left(T_\pi,T_a\right) s Y_\pi \Bigg], \label{eq:dTs_dT-final} \\
    \frac{d\,T_a(x)}{dx} =& - \frac{T_a}{x} +\frac{15}{2 \pi^2 H x  T_a^3}  \mathcal{T} \left(T_\pi,T_a\right) s^2 Y_\pi^2\,.\label{eq:dTa_dT-final}
\end{align}
Although we calculate the Hubble parameter by summing up  energy densities of all sectors, we do not further take into account the derivatives of evolving effective degrees of freedom in comoving energy densities.

\section{Energy transfer function}
\label{app:cross}

\begin{figure}[t]
    \centering
\includegraphics[width=0.5\textwidth]{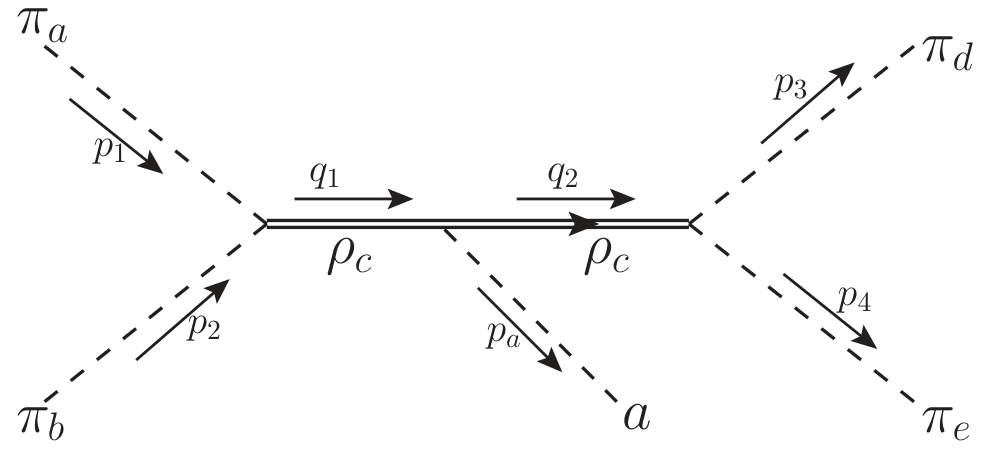}
    \caption{Feynman diagram for the process $2 \pi \rightarrow 2\pi+a$ mediated by $\rho$ mesons.}
    \label{fig:Fey_pi_pi_a_pi_pi}
\end{figure}

In this section, we present the detailed evaluation of the energy transfer function used in \autoref{sec:Bolt}. The kinetic equilibrium between DM and radiation is achieved by the $2 \pi \leftrightarrow  2\pi+a$ process, and the corresponding Feynman diagram is provided in~\autoref{fig:Fey_pi_pi_a_pi_pi}.

\subsection{Energy transfer function $\mathcal{Q}$ for $2\pi \leftrightarrow 2\pi+a $}

The process $2\pi \leftrightarrow   2\pi+a$ redistributes the energy from the hotter meson sector (pions) to colder radiation (axion), with the corresponding energy transfer function $\mathcal{Q}$ given by
\begin{align}
\mathcal{Q}=\int \prod \left[\frac{d^3 p}{(2\pi)^3 2 E} \right] (2\pi)^4 \delta^4\left(p_1+p_2-p_3-p_4-p_a \right) E_a |\mathcal{M}_{2\pi\leftrightarrow 2\pi+a}|^2 \left({\mathbb f}_{1} {\mathbb f}_{2}- {\mathbb f}_{3} {\mathbb f}_{4} {\mathbb f}_{a} \right)\,,
\label{eq:Q}
\end{align}
where ${\mathbb f}_i$ are distribution functions without taking into account the degrees of freedom, $E_a$ is the axion energy and $|\mathcal{M}_{2\pi\leftrightarrow 2\pi+a}|^2$ is the effective amplitude squared of all the $\pi_a + \pi_b \leftrightarrow \pi_a + \pi_b + a$ processes. 
Since $m_a \ll m_\pi$, the axion can be treated as effectively massless and is assumed to remain in thermal and chemical equilibrium with itself for the relevant timescales, leading to  ${\mathbb f}_{a}={\mathbb f}_{a}^{eq}(T_a)$. Consequently, the second term in the bracket of~\autoref{eq:Q} can be rewritten as 
\begin{align}
{\mathbb f}_{3} {\mathbb f}_{4} {\mathbb f}_{a}^{\rm{eq}}(T_a) = \frac{{\mathbb f}_{3} {\mathbb f}_{4}}{{\mathbb f}_{3}^{\rm{eq}} {\mathbb f}_{4}^{\rm{eq}}} \frac{{\mathbb f}_{a}^{\rm{eq}}(T_a)}{{\mathbb f}_{a}^{\rm{eq}}(T_\pi)} {\mathbb f}_{3}^{\rm{eq}}{\mathbb f}_{4}^{\rm{eq}} {\mathbb f}_{a}^{\rm{eq}}(T_\pi)\,.
\end{align}
Utilizing the principle of detailed balance following Ref.~\cite{Kolb:1990vq}, the equilibrium distribution functions satisfy $ {\mathbb f}_{3}^{\rm{eq}}(T_\pi) {\mathbb f}_{4}^{\rm{eq}}(T_\pi) {\mathbb f}_{a}^{\rm{eq}}(T_\pi) = {\mathbb f}_{1}^{\rm{eq}}(T_\pi) {\mathbb f}_{2}^{\rm{eq}}(T_\pi) $. In addition, since the ratio ${\mathbb f}_i/{\mathbb f}_i^{\rm{eq}} = n_i/n_i^{\rm{eq}}$ depends only on the temperature, it can be taken outside the integral in~\autoref{eq:Q}. That is, \autoref{eq:Q} is now given by
\begin{align}
\mathcal{Q}=\left(\frac{n_\pi}{n_\pi^{\rm{eq}}(T_\pi)}\right)^2 \int \prod \left[\frac{d^3 p}{(2\pi)^3 2 E} \right] (2\pi)^4 \delta^4\left(p_1+p_2-p_3-p_4-p_a \right) E_a |\mathcal{M}_{2\pi\leftrightarrow 2\pi+a}|^2 {\mathbb f}_{1}^{\rm{eq}}{\mathbb f}_{2}^{\rm{eq}} \left(1 - \frac{{\mathbb f}_{a}^{\rm{eq}}(T_a)}{{\mathbb f}_{a}^{\rm{eq}}(T_\pi)} \right)\,. \label{eq:Q1}
\end{align}

The energy transfer function $\mathcal{Q}$ is relevant primarily when the rate of the $2\pi \leftrightarrow 2\pi+a$ process exceeds the Hubble expansion rate, where $|T_\pi - T_a|/T_a \ll 1 $. Consequently, we can safely approximate the ratio ${\mathbb f}_{a}^{\rm{eq}}(T_a) / {\mathbb f}_{a}^{\rm{eq}} (T_\pi)$ to first order in the temperature difference $(T_\pi-T_a)$,
\begin{align}
\frac{{\mathbb f}_{a}^{\rm{eq}}(T_a)}{{\mathbb f}_{a}^{\rm{eq}}(T_\pi)} \approx 1 + E_a \frac{T_\pi-T_a}{T_\pi T_a} +\mathcal{O}[(T_\pi-T_a)^2]\,.
\end{align}
Because the pions belong to the same multiplet, they behave identically, so it suffices to consider a single representative process. We may therefore write
\begin{equation}
    |\mathcal{M}_{2\pi\leftrightarrow 2\pi+a}|^2
    = N_\pi(N_\pi-1)\,
    |\mathcal{M}_{\pi_1+\pi_3\leftrightarrow \pi_1+\pi_3+a}|^2 
\end{equation}
Eq.~\eqref{eq:Q1} then simplifies to
\begin{align}
\mathcal{Q}&= \left[\frac{N_\pi(N_\pi-1) (T_\pi-T_a)}{T_\pi T_a n_\pi^{\rm{eq}}(T_\pi)^2} \int \prod \left[\frac{d^3 p}{(2\pi)^3 2 E} \right] (2\pi)^4 \delta^4\left(p_1+p_2-p_3-p_4-p_a \right) E_a^2 |\mathcal{M}_{\pi_1 + \pi_3 \leftrightarrow \pi_1 + \pi_3 + a}|^2 {\mathbb f}_{1}^{\rm{eq}} {\mathbb f}_{2}^{\rm{eq}}\right] n_\pi^2  \label{eq:Q2} \\
&\equiv \mathcal{T}(T_\pi,T_a) n_\pi^2\,. \label{eq:funcdef}
\end{align}
To evaluate the integral in~\autoref{eq:Q2}, we expand the phase-space integral,
\begin{align}
I\equiv \int& \frac{d^3 p_1}{(2\pi)^3 2 E_1} \frac{d^3 p_2}{(2\pi)^3 2 E_2} {\mathbb f}_{1}^{\rm{eq}} {\mathbb f}_{2}^{\rm{eq}}  \nonumber\\
&\times \int \frac{d^3 p_3}{(2\pi)^3 2 E_3} \frac{d^3 p_4}{(2\pi)^3 2 E_4} \frac{d^3 p_a}{(2\pi)^3 2 E_a} (2\pi)^4 \delta^4\left(p_1+p_2-p_3-p_4-p_a \right) E_a^2 |\mathcal{M}_{\pi_1 + \pi_3 \rightarrow \pi_1 + \pi_3 + a}|^2\,. \label{eq:Qintegral}
\end{align}
For the second integral in~\autoref{eq:Qintegral}, all quantities except for $E_a$ are Lorentz invariant. As in the evaluation of the squared amplitude in~\autoref{sec:Meval}, we calculate the integral in~\autoref{eq:Qintegral} by working in both the centre-of-mass (CM) and lab frames. The value of $E_a$ in both frames are related by a Lorentz boost,
\begin{align}
E_a^{\rm{lab}} = \gamma ( E_a^{\rm{CM}} + \vec{p}_a^{\rm{CM}} \cdot \vec{\beta} ) = \gamma  ( E_a^{\rm{CM}} - \vec{q_2}^{\rm{CM}} \cdot \vec{\beta} )\,, \label{eq:EaLab}
\end{align}
where boost vector $\vec{\beta}$ and Lorentz factor $\gamma$ are given by, 
\begin{align}
\vec{\beta} = \frac{\vec{p}_+}{E_+}, \quad  \quad
\gamma = \frac{1}{\sqrt{1-\vec{\beta}^2}} = \frac{E_+}{\sqrt{s}}\,,
\end{align}
with $\vec{p}_+ = \vec{p}_1+\vec{p}_2$ and $E_+ = E_1+E_2$.

We parametrize the momenta $\vec{p}_+$ in spherical coordinates as,  $\vec{p}_+=|\vec{p}_+|((\sin(\theta_+)\cos(\phi_+),\sin(\theta_+)\sin(\phi_+),\cos(\theta_+))$, the~\autoref{eq:EaLab} can be rewritten as:
\begin{align}
E_a^{\rm{lab}} = \gamma  ( E_a^{\rm{CM}} - |\vec{q_2}^{\rm{CM}}| |\vec{\beta}| (\cos(\theta_q) \sin(\theta_+)\cos(\phi_+) +  \sin(\theta_q) \sin(\theta_+)\sin(\phi_+)) )\,.
\label{eq:EaLab1}
\end{align}
Since $d^3p_1 d^3p_2 = \frac{E_1 E_2}{4} dE_+ dE_- d\Omega_+ d\phi_1$, the second term in \autoref{eq:EaLab1} integrates to zero upon performing the integration over $\phi_+$. Consequently, for the evaluation of the phase-space integral, only the first term contributes, and \autoref{eq:EaLab1} simplifies to
\begin{align}
E_a^{lab} = \gamma  E_a^{CM} = \frac{E_+ (s-s_{34})}{2s}\,.
\end{align}

To evaluate the phase-space integral, we first combine the outgoing pion pair into an intermediate state, allowing the three-particle phase space to be factorized into two successive two-particle phase-space integrals:
\begin{align}
d\Phi_3 = \frac{1}{2\pi} \int ds_{34} d\Phi_2(s,s_{34},m_a^2) d\Phi_2(s_{34},m_\pi^2,m_\pi^2)\,. \label{eq:dphi3}
\end{align}
Since the two-particle phase-space element is Lorentz invariant, each integral can be evaluated in the most convenient reference frame. We evaluate the first phase-space integral, $d\Phi_2(s,s_{34},m_a^2)$ in the CM frame and the second, $d\Phi_2(s_{34},m_\pi^2,m_\pi^2)$ in the local frame. Consequently, \autoref{eq:dphi3} becomes
\begin{align}
d\Phi_3 = \frac{1}{2\pi} \frac{1}{(32\pi^2)^2} \int ds_{34} \int d\Omega_q \frac{\sqrt{\lambda(s,s_{34},m_a^2)}}{s} \int d\Omega_3^* \frac{\sqrt{\lambda(s_{34},m_\pi^2,m_\pi^2)}}{s_{34}}\,. 
\end{align}
After evaluating the second integral   (for outgoing momenta) in~\autoref{eq:Qintegral}, we obtain,
\begin{align}
I=\int& \frac{d^3 p_1}{(2\pi)^3 2 E_1} \frac{d^3 p_2}{(2\pi)^3 2 E_2} {\mathbb f}_{1}^{eq} {\mathbb f}_{2}^{eq} E_+^2 h(s)\,,\label{eq:Qintegral2}
\end{align}
where we have introduced 
\begin{align}
    h(s) \hat =  \int d\Phi_3 \left(\frac{s-s_{34}}{2s}\right)^2 |\mathcal{M}_{\pi_1 + \pi_3 \leftrightarrow \pi_1 + \pi_3 + a}|^2\,.
    \label{eq:hfunction}
\end{align}
Now the remaining integrand is a function only of $s$ and $E_+$. Following the procedure of Ref.~\cite{Gondolo:1990dk}, the integral in~\autoref{eq:Qintegral2} can be evaluated to obtain the $\mathcal{T}(T_\pi,T_a)$ in~\autoref{eq:funcdef}:
\begin{align}
\mathcal{T}(T_\pi,T_a) =  \frac{T_\pi-T_a}{T_\pi T_a n_\pi^{eq}(T_\pi)^2}  (2 \pi^2 T_\pi)\int ds s (s-4 m_\pi^2) h(s) \left(\sqrt{s} K_1\left(\frac{\sqrt{s}}{T_\pi}\right) + 3 T_\pi K_2\left(\frac{\sqrt{s}}{T_\pi}\right)\right)\,.
\end{align}

\subsection{Calculation of the Invariant Amplitude $\mathcal{M}$ for $2 \pi \leftrightarrow 2 \pi +a$}
\label{sec:Meval}

 The $\pi\pi\rho$ interaction arises from the chiral Lagrangian term in \autoref{eq:pipi_rho}, whereas the $a \rho \rho $ vertex receives contributions from both the chiral Lagrangian in \autoref{eq:chiral_rhorhoa} and the WZW term in \autoref{eq:WZW_rhorhoa}. Without loss of generality, we consider $\pi_1 +\pi_3 \to \rho_8$ and $\pi_1 +\pi_3 \to \rho_8+a$ as representative processes for our discussion. The Feynman rules for these interactions are given by
\begin{align}
\pi_1(p_1) \pi_3(p_2) \rightarrow \rho_8(q_1) :\quad\quad\quad & \frac{g_\rho}{\sqrt{2}} (p_2-p_1)^\mu, \label{eq:Feyn_ppr}\\
\rho_8^\mu(q_1) \rightarrow \rho_8^\nu(q_2) + a :\quad\quad\quad & i\left( 2 c_{\rho\rho}-\frac{K}{3\pi^2} \right)\frac{ g_\rho^2}{ f_a} \epsilon^{\mu\nu\alpha\beta} q_{2,\alpha}q_{1,\beta} \label{eq:Feyn_rra}. 
\end{align}
Using these Feynman rules , the invariant amplitude of the process $2 \pi \leftrightarrow 2\pi+a$ is given by
\begin{align}
i \mathcal{M} &=\left(\frac{g_\rho}{\sqrt{2}}(p_2-p_1)^\alpha\right) \frac{i\left(\eta_{\alpha\beta}-\frac{q_{1,\alpha}q_{1,\beta}}{m_\rho^2}\right)}{q_1^2-m_\rho^2+im_\rho \Gamma_\rho} \left(i \left( 2 C_{\rho\rho}-\frac{K}{3\pi^2} \right)\frac{ g_\rho^2}{ f_a} \epsilon^{\beta\gamma\sigma\delta} q_{2,\sigma}q_{1,\delta} \right) \frac{i\left(\eta_{\gamma\lambda}-\frac{q_{1,\gamma}q_{1,\lambda}}{m_\rho^2}\right)}{q_2^2-m_\rho^2+im_\rho \Gamma_\rho}\left(\frac{g_\rho}{\sqrt{2}}(p_3-p_4)^\lambda\right)\,.
\end{align}
Utilizing the antisymmetric property of the tensor $\epsilon^{\mu \nu \rho \sigma}$, we re-write the result   as 
\begin{align}
\mathcal{M} = -\left( 2 C_{\rho\rho}-\frac{K}{3\pi^2} \right)\frac{g_\rho^4}{f_a} \frac{1}{(q_1^2-m_\rho^2+im_\rho \Gamma)(q_2^2-m_\rho^2+im_\rho \Gamma)} \epsilon^{\alpha\beta\gamma\sigma} (p_{1}+p_{2})_\alpha (p_{1}-p_{2})_\beta p_{3,\gamma}p_{4,\sigma}\,. \label{eq:ampM1}
\end{align}
In the CM frame, where $(p_{1}+p_{2})_{\alpha}=(\sqrt{s},\vec{0})$, we perform an additional rotation to align the relative momentum $\vec{l}\equiv\vec{p}_{1}-\vec{p}_{2}$ with the $x$-axis. With this choice, the amplitude in~\autoref{eq:ampM1} simplifies in this frame to
\begin{align}
\mathcal{M} \propto \sqrt{s} \vec{l} \cdot (\vec{p}_3 \times \vec{p}_4) \propto 2 \sqrt{s} |\vec{p}_1| \cdot (\vec{p}_3 \times \vec{p}_4)_x\,,  \label{eq:ampM2} 
\end{align}
where $\vec{l} = \vec{p}_1- \vec{p}_2 = 2\vec{p}_1 = 2|\vec{p}_1| \hat{x}$ and the subscript $_x$ denotes the x-component of the vector $(\vec{p}_3 \times \vec{p}_4)$.

As shown in the previous section in \autoref{eq:hfunction}. In order to compute the energy transfer function we need to isolate the angular dependence of the amplitude in two different reference frames. For this purpose, we work in the CM frame of the outgoing pions $\vec{p}_3^*+\vec{p}_4^*=\vec{0}$, which we refer to as the local frame. This frame is obtained from the CM frame by applying a Lorentz boost along the direction $\vec{\beta}=-\vec{q}_2 / q_2^0$. Since Lorentz boosts leave the components of vectors transverse to the boost direction invariant, we have
\begin{align}
(\vec{p}_3 \times \vec{p}_4)= (\vec{p}_3 \times \vec{q}_2) = (\vec{p}_3^\perp \times \vec{q}_2) = (\vec{p}_3^{* \perp} \times \vec{q}_2) = (\vec{p}_3^* \times \vec{q}_2)\,,
\label{eq:vec_rel}
\end{align}
where $\perp$ denotes the component of the vector perpendicular to $\vec{q}_2$. We can parametrize the vectors $\vec{p}_3^*$ and $\vec{q}_2$ in terms of the angular coordinates as,
\begin{align}
    \vec{p}_3^* &= |\vec{p}_3^*|(\sin(\theta_3^*)\cos(\phi_3^*),\sin(\theta_3^*)\sin(\phi_3^*),\cos(\theta_3^*))\,, \\
    \vec{q}_2 &= |\vec{q}_2|(\cos(\theta_q),\sin(\theta_q),0)\,.
\end{align}
\autoref{eq:ampM2} can then be expressed using the angular coordinates as:
\begin{align}
\mathcal{M} \propto - 2 \sqrt{s} |\vec{p}_1| |\vec{p}_3^*| |\vec{q}_2| \cos(\theta_3^*) \sin(\theta_q)\,, \label{eq:ampM3}
\end{align}
where we have used the relation $(\vec{p}_3^{* \perp} \times \vec{q}_2)_x = -|\vec{p}_3^*| |\vec{q}_2| \cos(\theta_3^*) \sin(\theta_q)$.

Using~\autoref{eq:ampM3}, the squared amplitude $\left| \mathcal{M} \right|^2$ can be expressed in terms of Lorentz-invariant quantities as:
\begin{align}
|\mathcal{M}|^2 = \frac{1}{16} \left( C_{\rho\rho}+\frac {c_{gg}}{12\pi^2} \right)^2 \left( \frac{g_\rho^4 }{f_a}\right)^2 \frac{(s-4m_\pi^2)(s_{34}-4m_\pi^2)(s-s_{34})^2}{((s-m_\rho^2)^2+m_\rho^2 \Gamma^2)((s_{34}-m_\rho^2)^2 + m_\rho^2 \Gamma^2)} \cos(\theta_3^*)^2 \sin(\theta_q)^2\,. \label{eq:Msq}
\end{align}
In~\autoref{eq:Msq}, we have employed the following relations connecting the particle momenta and Lorentz invariant quantities:
\begin{align}
|\vec{p}_1| &= \frac{1}{2} \sqrt{\frac{\lambda(s,m_\pi^2,m_\pi^2)}{s}} = \frac{1}{2} \sqrt{s-4m_\pi^2}\,,\\
|\vec{p}_3^*| &= \frac{1}{2} \sqrt{\frac{\lambda(s_{34},m_\pi^2,m_\pi^2)}{s_{34}}}=\frac{1}{2} \sqrt{s_{34}-4m_\pi^2}\,,\\
|\vec{q}_2| &= \frac{1}{2} \sqrt{\frac{\lambda(s,s_{34},m_a^2)}{s}} = \frac{1}{2\sqrt{s}} (s-s_{34})\,,
\end{align}
where the Källén function $\lambda(s,m_a^2,m_b^2)=(s-(m_a+m_b)^2)(s-(m_a-m_b)^2)$ and $s_{34}=(p_3+p_4)^2$.

As already mentioned in the main text, for numerical simulations of DM freeze-out, we always take the non-relativistic limit of interaction rates. That is, in practice we only include energy/velocity-dependence in the numerator of \autoref{eq:Msq}, while setting $s\approx s_{34}\approx 4 m_\pi^2$ in its denominator. 
In addition, later we adopt numerically $m_\rho = 2.5m_\pi$, which is far from the pole of the propagator, so the $\rho$-decay width, $\Gamma_\rho$, is safely set to vanish in the cross section above.

\section{Extended Numerical Results}
\label{sec:appendix_d}

In the section, we investigate in more detail the dependence of dark pion freeze-out on the relevant parameters, such as $m_\pi/f_\pi$, $m_\pi$, and $f_a$, as well as provide its final relic abundance for more parameter choices.

\subsection{Dependence of the relic abundance on the ratio, ${m_\pi}/{f_\pi}$}

\begin{figure}[t]
    \centering
    \includegraphics[width=0.48\linewidth]{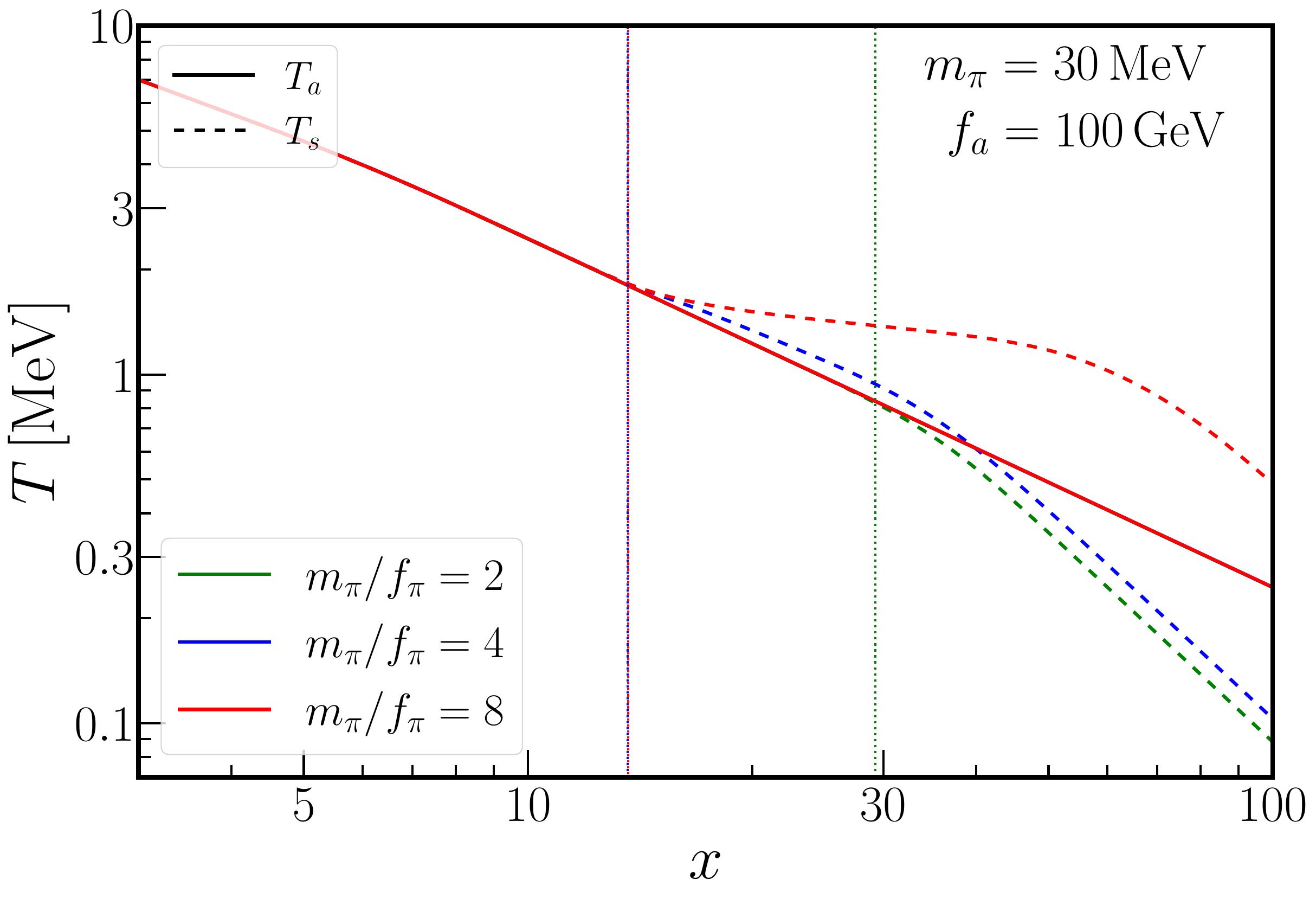}
    \includegraphics[width=0.495\linewidth]{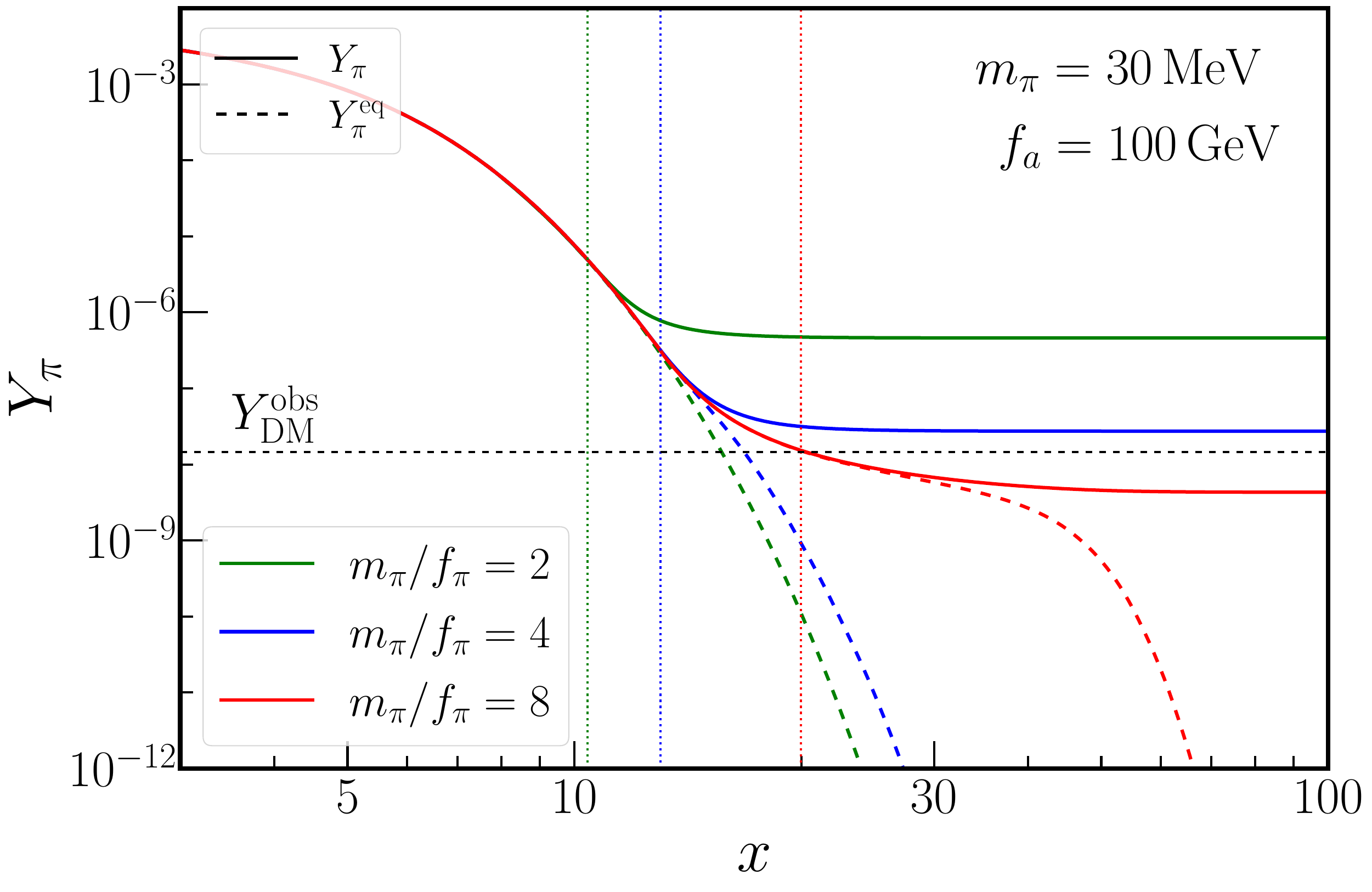}
    \caption{Temperature evolution (left panel) and pion yield (right panel) as functions of $x = m_\pi/T_\gamma$ for   $m_\pi = 30\text{ MeV}$ and  $f_a = 10^2\text{ GeV}$, shown for several values of  $m_\pi/f_\pi$. Vertical lines mark kinetic decoupling ($x_k$) in the left panel and chemical decoupling ($x_f$) in the right panel. For $m_\pi/f_\pi = 2$, kinetic decoupling occurs well after freeze-out, meaning the yield is controlled strictly by the number-changing process. For $m_\pi/f_\pi = 4$, the system is borderline ($x_k \simeq x_f$), but kinetic decoupling still has little impact on the final yield. For $m_\pi/f_\pi = 8$, the kinetic-decoupling point has saturated, and freeze-out is delayed by the intervening slow-cooling cannibal phase, leading to a larger yield than expected from the cross-section scaling alone.}
    \label{fig:yield_fpi}
\end{figure}

The effective coupling within the meson sector, $m_\pi / f_\pi$, mostly affects the relic abundance by modifying the self-annihilation rate of dark pions. Quantitatively, the $3\pi \to 2\pi$ annihilation rate scales as $\langle \sigma v^2 \rangle_{3\pi\to2\pi} \propto \left(m_\pi / f_\pi \right)^{10}$. Therefore, increasing  $m_\pi / f_\pi$  enhances this number-changing process, which tends to lower the final pion yield. As shown in~\autoref{fig:yield_fpi}, larger values of $m_\pi / f_\pi$ correspond to smaller final relic abundances of $Y_\pi$.

Interestingly, during freeze-out a very large $m_\pi / f_\pi$ may lead to  too much energy injection into the meson sector, more than that the inelastic scattering $2\pi \leftrightarrow 2\pi + a$ can dissipate into the axion bath. As a result, $T_s/T_a$ (or similarly,  $T_s/T_\gamma$) increases, partially canceling the effect of an enhanced self-annihilation. In the extreme case where kinetic decoupling occurs before chemical freeze-out ($x_k < x_f$),  the relic abundance becomes less and less sensitive to $m_\pi / f_\pi$. That is, for $x_k \ll x_f$,  the final relic abundance becomes insensitive to the value of $m_\pi / f_\pi$.

\begin{figure}[t]
    \centering
    \includegraphics[width=0.48\linewidth]{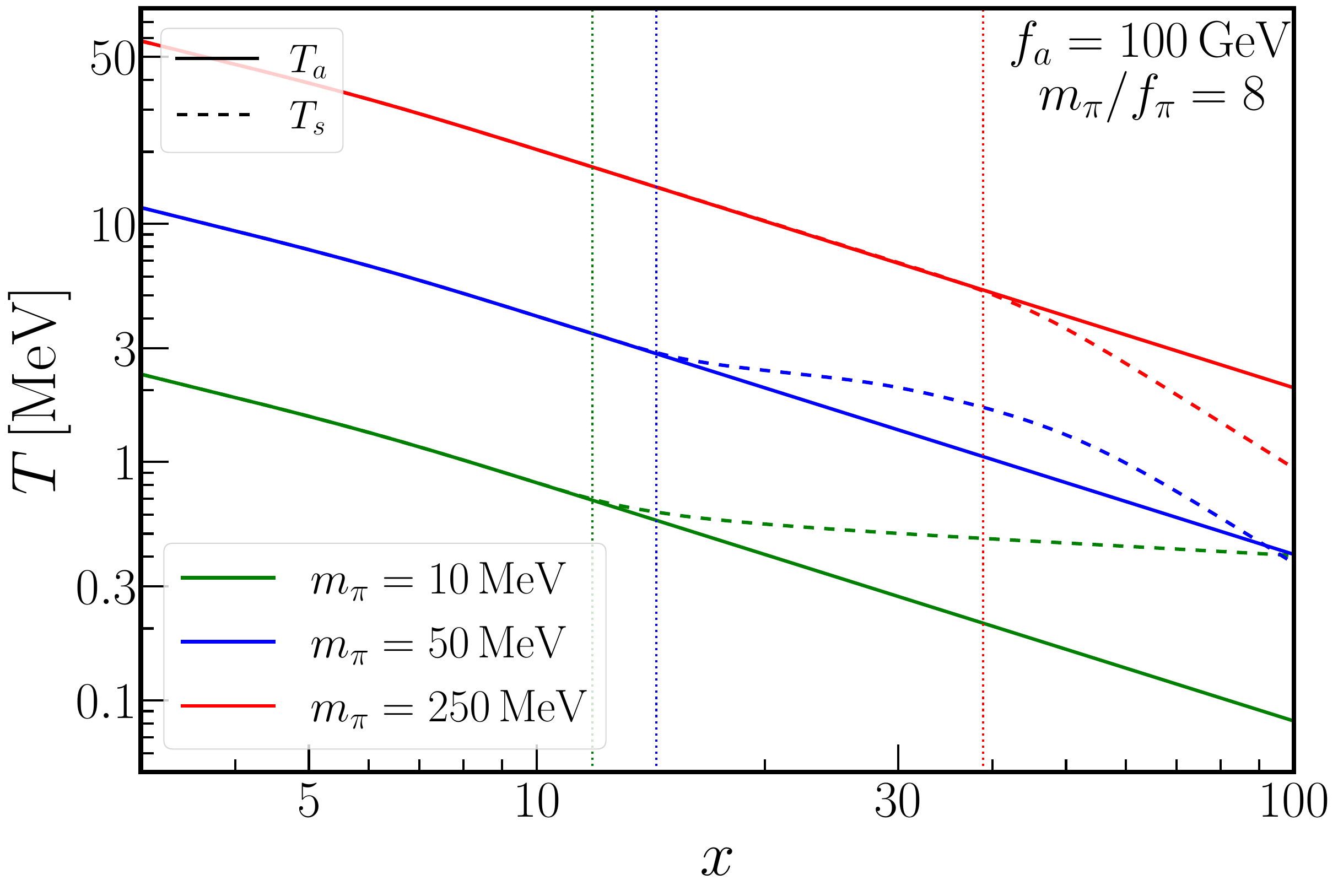}
    \includegraphics[width=0.495\linewidth]{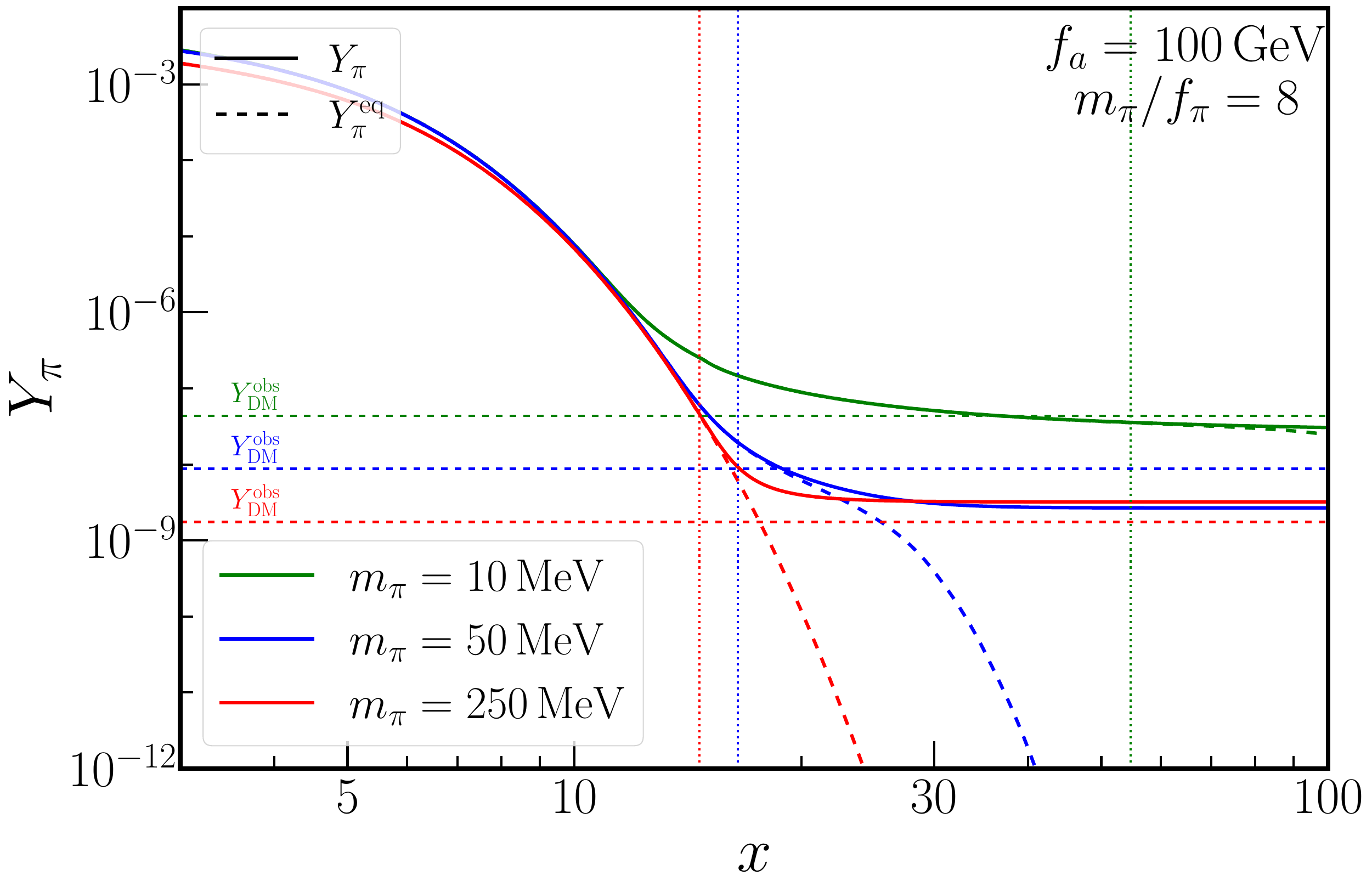}
    \caption{Temperature evolution (left panel) and pion yield (right panel) as functions of $x=m_\pi/T_\gamma$ for fixed $m_\pi/f_\pi=8$ and $f_a=10^2$ GeV, shown for several values of $m_\pi$. Vertical lines mark kinetic decoupling ($x_k$) in the left panel and chemical decoupling ($x_f$) in the right panel. Unlike $f_\pi$, which mostly affects the $3\pi\to2\pi$ rate, $m_\pi$ also controls the strength of the energy transfer function $\mathcal{Q}$, making its impact on the thermal evolution more pronounced. Consequently, lowering $m_\pi$ leads to earlier kinetic decoupling and a stronger cannibal phase, which may enhance the final yield, as the case of $m_\pi=10$ MeV.}
    \label{fig:mpi_dependence}
\end{figure}

\subsection{Dependence of the relic abundance on the DM mass, $m_\pi$}

The dependence on the dimensional parameter, $m_\pi$, is trickier. For self-annihilating processes,  the DM mass enters the $3\pi\to2\pi$ annihilation rate as $\langle\sigma v^2\rangle_{3\pi\to2\pi}\propto m_\pi^{-7}$ for a fixed value of  $m_\pi/f_\pi$. Lowering $m_\pi$ thus significantly enhances the annihilation rate, affecting both DM number reduction and   temperature evolution, in analogy with the effect of increasing the ratio $m_\pi/f_\pi$  in the previous subsection. Therefore, for the parameter region where $x_k\gtrsim x_f$, the exact time of kinetic decoupling is irrelevant for freeze-out, and smaller values of $m_\pi$ yield less final relic abundances. 

Regarding the internal Bremsstrahlung process,  which maintains kinetic equilibrium between the meson and axion sectors,  for a fixed $f_a$ smaller $m_\pi$ reduces  the energy transfer efficiency. This causes kinetic decoupling to occur earlier,  impeding further dark pion number-reduction. A smaller $m_\pi$ also increases the $3\pi\to 2\pi$ cross section, so the two effects therefore compete, and the net 
annihilation efficiency depends on which one dominates. 
Deep in the parameter region of $x_k < x_f$,   smaller $m_\pi$ typically lead to  larger relic abundance. This behavior is illustrated in~\autoref{fig:mpi_dependence} for $m_\pi/f_\pi=8$ and $f_a=10^2$ GeV, where the case with $m_\pi=10$ MeV yields a larger abundance than those with heavier pions. This can also be observed in~\autoref{fig:param_scan_fa}, where on the left part of each panel,  the final relic abundance should increase with smaller values of $m_\pi$ for $x_k < x_f$.

Note that the dependence on $m_\pi$ would be different if one instead fixes both dimensionless parameters, $m_\pi/f_\pi$ and $m_\pi/f_a$. In this case, reducing  $m_\pi$ increases relevant cross sections and thus enhance both the efficiencies of self-annihilation and internal Bremsstrahlung processes,  resulting in a smaller final relic abundance.

\subsection{Impact of the Peccei-Quinn scale, $f_a$, on the viable parameter space}

At last, we briefly discuss the impact of the PQ   scale $f_a$ on the relic abundance. In addition, we  provide a scan on  the ($m_\pi$,\,$m_\pi/f_\pi$) parameter plane for another two values of $f_a$,  serving as a complement to the scan plot in the main text.
Essentially, lowering   $f_a$ enhances the cross-section for the $2\pi \leftrightarrow 2\pi + a$ interaction,  enlarging the parameter space where kinetic equilibrium with pions and the axion bath can be maintained until DM freeze-out. 

\begin{figure}[t]
    \centering
    \includegraphics[width=0.49\textwidth]{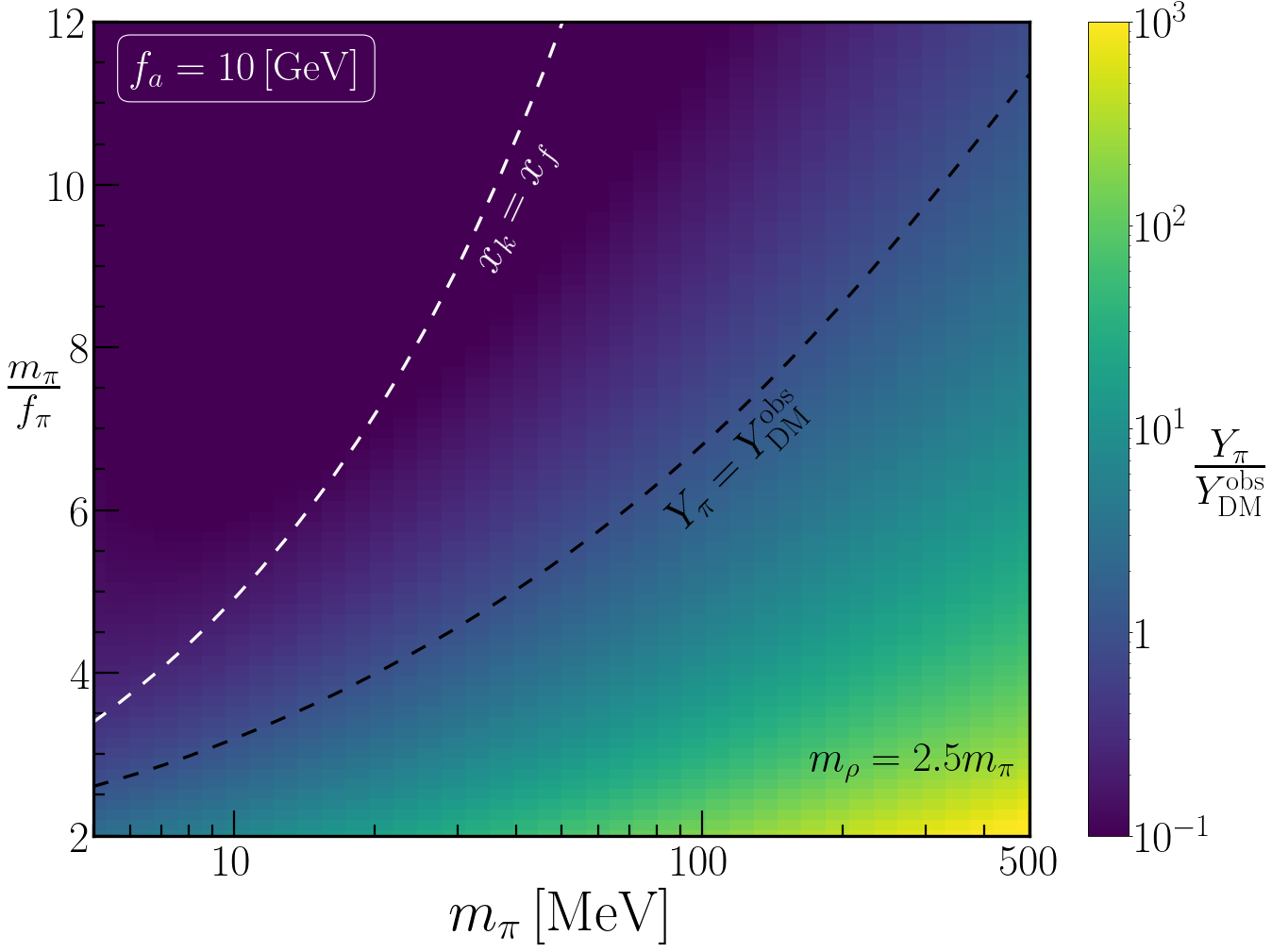}
    \includegraphics[width=0.49\textwidth]{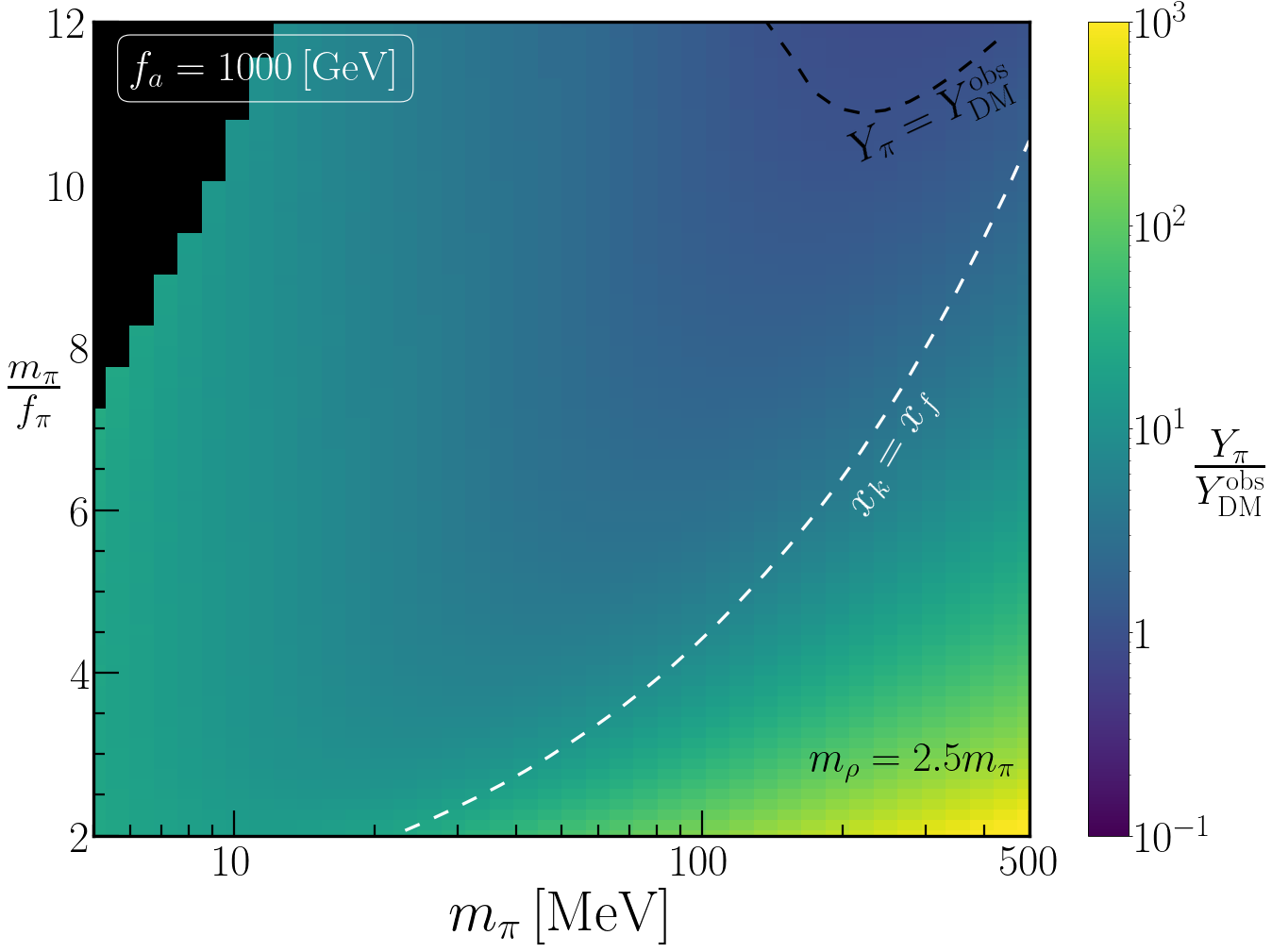}
    \caption{Parameter scan in the ($m_\pi$,\,$m_\pi/f_\pi$) plane for $f_a = 10\text{ GeV}$ (left) and $f_a = 10^3\text{ GeV}$ (right). The white dashed line corresponds to $x_k = x_f$, where kinetic and chemical decoupling coincide. The black dashed line denotes the contour that reproduces the observed DM relic abundance. In the dark shaded region, chemical freeze-out does not occur before $x = 500$.}
    \label{fig:param_scan_fa}
\end{figure}

In~\autoref{fig:param_scan_fa}, we show the final relic abundance  for $f_a=10\,\rm{GeV}$ (left panel) and $f_a=10^3\,\rm{GeV}$ (right). For $f_a=10\,\rm{GeV}$, kinetic energy transfer between pions and axions is generally efficient. Consequently, in most of the parameter space, kinetic decoupling occurs after freeze-out ($x_k>x_f$), and the freeze-out mechanism is analogous to the standard SIMP mechanism. Conversely, for $f_a=10^3\,\rm{GeV}$, the $2\pi \leftrightarrow 2\pi + a$ process is very suppressed, leading to earlier kinetic decoupling ($x_k < x_f$). Consequently, the resulting cannibal phase traps heat in the meson sector, which systematically increases the relic abundance and risks overproducing the DM relic abundance. 
Note that there is a significant parameter region at the top-left corner  where the freeze-out does not complete before $x=500$. We do not compute this region further and mark it in black.

\end{document}